\documentclass{ieeeaccess}
\usepackage{cite}
\usepackage{url}
\usepackage{threeparttable}
\usepackage{booktabs}
\usepackage{siunitx}
\usepackage{adjustbox}
\usepackage{amsmath,amssymb,amsfonts}
\usepackage{algorithmic}
\usepackage{graphicx}
\usepackage[normalem]{ulem}
\usepackage{textcomp}
\usepackage{underscore}
\usepackage{hyperref}

\def\BibTeX{{\rm B\kern-.05em{%
\sc i\kern-.025em b}\kern-.08em
T\kern-.1667em\lower.7ex\hbox{E}\kern-.125emX}}

\begin{document}
\history{Date of publication xxxx 00, 0000, date of current version xxxx 00, 0000.}
\doi{10.1109/ACCESS.2017.DOI}

\title{Evaluating Generalization and Robustness in Russian Anti-Spoofing: The RuASD Initiative}
\author{\uppercase{Ksenia Lysikova}\authorrefmark{1},
\uppercase{Kirill Borodin}\authorrefmark{1}, and
\uppercase{Grach Mkrtchian}\authorrefmark{1}}
\address[1]{Moscow Technical University of Communications and Informatics, Moscow, Russia, 111038}

\markboth
{Author \headeretal: Preparation of Papers for IEEE TRANSACTIONS and JOURNALS}
{Author \headeretal: Preparation of Papers for IEEE TRANSACTIONS and JOURNALS}

\corresp{Corresponding author: Kirill Borodin (e-mail: k.n.borodin@mtuci.ru).}

\begin{abstract}
RuASD (Russian AntiSpoofing Dataset) is a dedicated, reproducible benchmark for Russian-language speech anti-spoofing designed to evaluate both in-domain discrimination and robustness to deployment-style distribution shifts. It combines a large spoof subset synthesized using 37 modern Russian-capable TTS and voice-cloning systems with a bona fide subset curated from multiple heterogeneous open Russian speech corpora, enabling systematic evaluation across diverse data sources. To emulate typical dissemination and channel effects in a controlled and reproducible manner, RuASD includes configurable simulations of platform and transmission distortions, including room reverberation, additive noise/music, and a range of speech-codec transcodings implemented via a unified processing chain. We benchmark a diverse set of publicly available anti-spoofing countermeasures spanning lightweight supervised architectures, graph-attention models, SSL-based detectors, and large-scale pretrained systems, and report reference results on both clean and simulated conditions to characterize robustness under realistic perturbation pipelines. The dataset is publickly available at \href{https://huggingface.co/datasets/MTUCI/RuASD}{\underline{Hugging Face}} and \href{https://modelscope.cn/datasets/lab260/RuASD}{\underline{ModelScope}}.
\end{abstract}


\titlepgskip=-15pt

\maketitle

\section{Introduction}
\label{sec:introduction}
Neural speech synthesis (TTS) and voice conversion (VC) have advanced rapidly in recent years, enabling highly realistic speech generation and raising concerns about audio disinformation and attacks on voice-enabled applications \cite{tan2021_neural_tts_survey,shue2023_styletts2,asvspoof2021_accelerating,asvspoof2021_summary}.
This shift is reflected in ASVspoof~2021, which introduced the DeepFake (DF) track to evaluate detection of manipulated speech distributed ``posted online'' under compression and channel distortions \cite{asvspoof2021_accelerating,asvspoof2021_summary}.
Despite improved robustness to encoding and transmission effects, DF conditions remain difficult due to limited generalization across data sources and generation settings \cite{asvspoof2021_summary}.
Therefore, anti-spoofing evaluation should assess both in-domain discrimination performance and robustness to distribution shifts caused by diverse generators and post-processing pipelines, including transcoding \cite{asvspoof2021_accelerating,asvspoof2021_summary}.

Adversarial pipelines often diverge from detector training conditions: new generative models and evolving artifacts, combined with variability in dissemination channels and compression, can substantially alter observable spoofing cues \cite{asvspoof2021_accelerating,asvspoof2021_summary}.
Robust transfer under such out-of-domain shifts is therefore a core objective of modern benchmarks, including ASVspoof~2021, which adopts a mismatched training/development setup to better approximate deployment scenarios where attack characteristics are not known in advance \cite{asvspoof2021_accelerating}.
Complementing controlled protocols, recent ``in-the-wild'' datasets further assess generalization across diverse real-world sources and channel conditions \cite{unmasking_real_world_audio_deepfakes}.

In parallel, significant progress has been made in the development of multilingual resources aimed at systematic assessment of cross-linguistic generalization.
Specifically, MLAAD explicitly addresses the bias of existing datasets toward a limited number of languages and proposes a multilingual corpus, including Russian, for training and evaluating anti-spoofing models under conditions of linguistic diversity.
MLAAD also demonstrates the practical motivation for multilingual coverage, linking it to improved cross-dataset transferability of detectors\cite{MLAAD}.

Nevertheless, a gap persists for the Russian language between the rapid advancement of Russian TTS solutions and the availability of benchmarking resources that enable representative and reproducible evaluation of countermeasures under modern Russian-language threat models.
Given that key limitations of DF scenarios are associated with generalization to novel generation sources and variable channel conditions, a Russian-focused corpus and unified reproducible protocol are required to complement existing competitions and multilingual initiatives, thereby ensuring proper assessment of detector transferability in the Russian-language.

To address the lack of a dedicated, reproducible corpus for Russian-language speech anti-spoofing under simulation of realistic channel and platform processing conditions, we introduce \textbf{RuASD} (Russian AntiSpoofing Dataset) and make the following contributions:

\begin{itemize}
  \item \textbf{RuASD: A Russian-language anti-spoofing dataset.}
  We introduce RuASD, combining bona fide speech collected from open sources with spoofed speech synthesized using a diverse set of modern Russian-capable TTS systems (including both open-source models and commercial APIs).

  \item \textbf{Reproducible simulation of real-world dissemination.}
  We provide controlled, configurable augmentations that emulate typical degradation pipelines encountered in practice, including (i) room reverberation using RIRs (RIRS), (ii) additive noise and music using MUSAN, and (iii) codec-based degradations implemented via a unified processing chain. 

  \item \textbf{Evaluation of modern open source detectors.}
  We benchmark a diverse set of publicly available anti-spoofing detectors \textbf{RuASD}, covering both task-specific supervised architectures and large-scale pretrained/SSL-based systems, to establish reproducible reference results under clean conditions and controlled simulations of degraded-channel conditions.
\end{itemize}

\section{Related Work}

\subsection{Speech Anti-Spoofing Datasets and Benchmarks}
\begin{table}[t]
\centering
\caption{Compact comparison of representative audio anti-spoofing datasets. ``Models'' denotes the number of spoof generation systems when reported. ``Utterances'' is the number of audio samples/segments reported by the dataset authors.}
\label{tab:datasets_compact}
\setlength{\tabcolsep}{2.6pt}
\renewcommand{\arraystretch}{1.05}

\begin{tabular}{lcccc}
\toprule
\textbf{Dataset} & \textbf{Models} & \textbf{Russian models} & \textbf{Utterances} & \textbf{Hours} \\
\midrule
ASVspoof 2019 LA \cite{asvspoof2019_wang}        & 19   & 0 & 121{,}461            & 109.65 \\
ASVspoof 2021 LA \cite{asvspoof2021_yamagishi}  & 13   & 0 & 164{,}612            & 132.51 \\
ASVspoof 2021 DF \cite{asvspoof2021_yamagishi}  & 100+ & 0 & \uline{593{,}253}    & 507.73 \\
FakeOrReal (FoR) \cite{reimao2019_forr}         & 7    & 0 & 195{,}541            & 72.80 \\
WaveFake \cite{frank2021_wavefake}                        & 9    & 0 & 136{,}085            & 216.21 \\
In-The-Wild \cite{muller2022_in_the_wild}        & --   & 0 & 31{,}779             & 38 \\
ADD 2022 \cite{yi2022_add}                      & --   & 0 & 493{,}123            & 370.47 \\
SpeechFake-MD~\cite{huang2025speechfake}                                   & 40   & 1  & \textbf{1{,}335{,}492} & \textbf{4855} \\
SynHate~\cite{ranjan25_interspeech}                                         & 1    & 1  & --                  & 271.70 \\
XMAD-Bench\cite{Ciobanu2025xmad}                                      & 17   & 2  & 102{,}146           & 193.25 \\
MLAAD \cite{MLAAD}                              & 140  & \uline{9}  & 298{,}000           & 687.40 \\
\textbf{RuASD}                                   & 37   & \textbf{37} & 375{,}363 & \uline{925.70} \\
\bottomrule
\end{tabular}
\end{table}

Speech anti-spoofing evaluation is largely shaped by public benchmarks that define standardized partitions, attack taxonomies, and reporting protocols, with the ASVspoof series being the most widely used reference point. \cite{asvspoof2019_wang,asvspoof2021_yamagishi}
ASVspoof covers multiple threat models, including logical access (TTS/VC) and physical access (replay), and ASVspoof~2021 additionally introduced the DeepFake (DF) track to better reflect ``posted online'' distribution effects. \cite{asvspoof2021_yamagishi}

Recent benchmarks emphasize increased realism and scale.
ASVspoof~5 is based on crowdsourced speech collected under diverse acoustic conditions and expands speaker coverage to approximately 2{,}000 speakers; it also includes attacks generated by 32 algorithms (including adversarial attacks) and defines seven speaker-disjoint partitions to support robust evaluation. \cite{asvspoof5_wang2025}

Beyond the ASVspoof family, existing datasets can be grouped into three complementary directions.
(i) \emph{In-the-wild} corpora such as In-The-Wild and FakeOrReal aim to stress-test detectors under uncontrolled collection and dissemination pipelines, where generators and post-processing chains are not fully known. \cite{muller2022_in_the_wild,reimao2019_forr}
(ii) \emph{Multilingual} resources such as MLAAD address language coverage by providing large-scale synthesized speech (687.4 hours in 51 languages generated by 140 TTS systems), enabling cross-lingual transfer studies and reducing the English-centric bias of many benchmarks. \cite{MLAAD}
(iii) \emph{Controlled synthetic} corpora such as WaveFake focus on analyzing model- and vocoder-specific artifacts and are frequently used for auxiliary evaluation and transfer experiments. \cite{frank2021_wavefake}

Table~\ref{tab:datasets_compact} summarizes representative datasets and contrasts them with \textbf{RuASD}.
While prior benchmarks either focus on standardized protocols (ASVspoof), emphasize uncontrolled real-world collection (in-the-wild), or prioritize language breadth (multilingual datasets), \textbf{RuASD} is designed to provide \emph{depth for Russian} under a reproducible robustness protocol.
In particular, RuASD combines (a) spoof speech generated by 37 modern Russian-capable TTS systems with (b) controlled, deployment-inspired channel/codec perturbations, enabling systematic evaluation of both in-domain discrimination and robustness under realistic distribution shift.

\subsection{Anti-Spoofing Architectures and Methods}
Modern countermeasures are typically implemented as supervised binary classifiers (bona fide vs.~spoof) \cite{asvspoof2021_yamagishi}, differing mainly in the input representation and the modeling of spectro-temporal artifacts \cite{aasist_2021_jung}.
For clarity and comparability, recent work commonly groups approaches by architectural bias and the source of prior knowledge (task-specific supervised learning vs.~pretraining) \cite{MLAAD,ssl_wav2vec2_tak2022}.

\textit{Convolutional and temporal modeling backbones.}
A large fraction of competitive systems uses convolutional feature extractors coupled with explicit temporal modeling to capture local and mid-range artifact cues while remaining computationally efficient \cite{asvspoof2019_wang}.
Res2TCNGuard and ResCapsGuard are representative examples of lightweight supervised backbones tailored for spoofing detection \cite{borodin2024_rescaps_res2tcn}.
TCM-ADD proposes temporal--channel modeling within multi-head self-attention in an SSL-based Conformer pipeline for synthetic speech detection \cite{truong2024_tcmadd}.

\textit{Graph-attention models for spectro-temporal relations.}
Graph-based anti-spoofing models explicitly model relationships between spectral and temporal patterns via attention over structured node representations, aiming to capture long-range dependencies and subtle cross-time/frequency correlations \cite{aasist_2021_jung}.
A prominent line of work is AASIST and its extensions \cite{aasist_2021_jung,viakhirev2025_scalable_aasist}; AASIST3 is a recent representative extension that enhances AASIST-style modeling \cite{aasist3_2024_borodin}.
Recent work also studies component-level interpretability of multi-branch anti-spoofing architectures such as AASIST3, linking branch-level contribution patterns to per-attack reliability and exposing confidently-wrong failure modes (``flawed specialization'') \cite{viakhirev2026_interpreting_multibranch}.

\textit{SSL-based and large-scale pretrained detectors.}
A recent trend is to leverage pretrained speech encoders as front-ends to obtain higher-level representations that improve transferability under unseen generation sources and post-processing pipelines, including channel and codec distortions~\cite{ssl_wav2vec2_tak2022,MLAAD}.
In this paradigm, a self-supervised speech encoder (e.g., wav2vec~2.0 or XLS-R) is used as a feature extractor and fine-tuned for spoofing detection with a task-specific back-end that aggregates frame-level representations and produces bona fide/spoof decisions~\cite{ssl_wav2vec2_tak2022}.
wav2vec~2.0 anti-spoofing studies the use of wav2vec~2.0 representations for audio deepfake detection and analyzes their impact on cross-dataset generalization~\cite{ssl_wav2vec2_tak2022}.
SLS with XLS-R leverages XLS-R self-supervised representations and applies sensitive layer selection for deepfake speech detection~\cite{zhang2024_sls_mm}.
Nes2Net targets foundation-model-driven setups by introducing a lightweight nested back-end designed to process high-dimensional representations without explicit dimensionality reduction layers~\cite{liu2025_nes2net}.
In parallel, Arena-1B and Arena-500M provide large-capacity universal anti-spoofing checkpoints; the released models use a Conformer-based backbone in the provided implementation and are distributed via public model cards~\cite{kulkarni2025_df_arena_1b_hf,kulkarni2025_df_arena_500m_hf}.

\textit{Implication for Russian-focused evaluation.}
Overall, existing benchmarks emphasize either standardized English-centric protocols (ASVspoof family) \cite{asvspoof2019_wang,asvspoof2021_yamagishi,asvspoof5_wang2025},
uncontrolled ``in-the-wild'' stress tests \cite{muller2022_in_the_wild,reimao2019_forr},
or multilingual breadth (MLAAD) \cite{MLAAD}.
These complementary designs motivate a focused Russian-language evaluation setup that combines (i) modern Russian-capable TTS sources and (ii) controlled, reproducible channel/codec perturbations, while reporting baselines spanning supervised CNN/TCN-style models, graph-attention detectors, and large-scale SSL-based approaches \cite{asvspoof5_wang2025,MLAAD,ssl_wav2vec2_tak2022}.

\section{Dataset Construction}
\textbf{RuASD} is designed to stress-test Russian-language anti-spoofing countermeasures under two complementary sources of distribution shift: (i) variation across spoof generators (diverse modern Russian-capable TTS and voice-cloning systems) and (ii) variation introduced by controlled simulations of dissemination and channel effects (noise, reverberation, and transcoding).
To keep the evaluation reproducible and interpretable, we explicitly control generator coverage in the spoof subset and source diversity in the bona fide subset, and we provide a deterministic selection protocol for data curation.

Within \textbf{RuASD}, spoof utterances were collected using multiple Russian text-to-speech (TTS) systems described in Sec.~\ref{sec:tts_models} and synthesized following the protocol in Sec.~\ref{sec:spoof_protocol}. Bona fide utterances were collected from 10 different Russian speech corpora and selected according to the protocol in Sec.~\ref{sec:bonafide_protocol}; the data sources are summarized in Sec.~\ref{sec:bonafide_sources}.
\subsection{TTS models}
\label{sec:tts_models}
To promote generalization-oriented evaluation, we include TTS systems spanning multiple synthesis paradigms and deployment modalities: open-source neural TTS checkpoints, voice-cloning pipelines, classical/offline engines, and commercial cloud APIs.
This design aims to avoid over-reliance on artifacts specific to a single model family and to provide a more realistic approximation of evolving Russian-language spoofing threats.

The ESpeech models include the ESpeech-TTS-1 checkpoints SFT-95k, SFT-256k, as well as their RL versions (RL-V1 and RL-V2), and the Podcaster checkpoint (SFT). All models are implemented based on the F5‑TTS architecture~\cite{chen2024f5tts}. The Podcaster model is trained exclusively on podcasts and focuses on more natural, spontaneous speech, while RL-V1 is based on SFT-95k and RL-V2 on SFT-256k. Two additional Russian-specific F5 fine-tunes are used: one trained on 5,000 hours of mixed Russian/English speech with explicit stress control and another adapted for 813k steps on a mixed Russian/English training setup.

The VITS family~\cite{kim2021vits} is represented by multiple end-to-end systems: a dedicated Russian checkpoint, PiperTTS (speakers \textit{denis}, \textit{dmitri}, \textit{irina}, \textit{ruslan}), TeraTTS (speakers \textit{natasha}, \textit{glados}, \textit{glados2}), and the Massively Multilingual Speech (MMS) Russian model~\cite{pratap2023mms}. This category also includes two multi-speaker checkpoints that synthesize speech directly from text with automatic stress placement, as well as a single-stage VITS2 model fine-tuned on the Natasha dataset~\cite{kong2023vits2}.

GPT-SoVITS implements a voice-cloning pipeline where a GPT module predicts intermediate speech representations from text, which are then reconstructed into waveforms by a SoVITS decoder~\cite{gptsovits2024}.

The multilingual voice-cloning model XTTS-v2~\cite{casanova2024xtts} encodes speaker conditioning from a reference utterance into fixed-length latents to condition a GPT-2–style discrete-code generator, subsequently reconstructing the waveform via a HiFi-GAN vocoder~\cite{casanova2024xtts,kong2020hifigan}.
In addition to the base checkpoint, two Russian fine-tunes of XTTS-v2 are employed.The first checkpoint is trained on approximately 40 hours of speech.
The second checkpoint employs an auxiliary IPA-and-stress front-end (a statistical IPA transcription module and two BERT-based stress models); the author reports training the stress models on approximately 3GB of stress-marked text and fine-tuning the acoustic model on approximately 60 hours of speech from RUSLAN and Russian Common Voice \cite{casanova2024xtts}.

Several Russian TTS pipelines employ FastPitch as the acoustic model\cite{lancucki2021fastpitch}. Across all FastPitch-based pipelines, the acoustic model predicts a mel-spectrogram from text, and a neural vocoder subsequently synthesizes the waveform \cite{lancucki2021fastpitch}.
In our setup, both the FastPitch acoustic model and the corresponding vocoder are trained on the RUSLAN corpus.

The first pipeline uses a Russian FastPitch checkpoint together with a BERT-based IPA phonemizer and a HiFi-GAN vocoder\footnote{\url{https://huggingface.co/bene-ges/tts_ru_hifigan_ruslan}} \cite{kong2020hifigan}.

The second pipeline consisted of a grapheme-to-phoneme converter, whose output is provided to the same FastPitch checkpoint, followed by the same HiFi-GAN checkpoint~\cite{kong2020hifigan}.

The third pipeline is RussianFastPitch, which combines FastPitch and the WaveGlow vocoder \cite{lancucki2021fastpitch,prenger2019waveglow}.
It additionally applies prosody post-processing by modifying the fundamental frequency contour to enforce one of six intonation constructions, enabling explicit intonation control.

Bark~\cite{suno2023bark} is a transformer-based TTS model that synthesizes speech via discrete tokens.
It first predicts \textit{semantic tokens} that represent high-level linguistic content, then converts them into \textit{coarse tokens} (the first two EnCodec codebooks) and then predicts \textit{fine tokens} (the remaining EnCodec codebooks) to represent high-resolution acoustic detail. The predicted codebooks are finally decoded into a waveform using an EnCodec decoder.

Grad-TTS is a diffusion-based acoustic model that synthesizes mel-spectrograms via a score-based decoder with MAS-based text–acoustic alignment. \cite{popov2021gradtts}
Inference speed and quality are controlled by the number of reverse diffusion steps, and the waveform is produced using a separate neural vocoder. \cite{popov2021gradtts}

Fish-Speech is a non-G2P TTS system that formulates synthesis as the generation of discrete audio tokens\cite{liao2024fishspeech}.
The input text is processed by an LLM, after which the core model predicts audio-codec tokens using a fast-slow Dual-AR scheme; waveform is then reconstructed by a dedicated decoder.
As the discrete representation, it uses GFSQ (Grouped Finite Scalar Vector Quantization), which encodes audio into codebook indices, and employs Firefly-GAN for waveform reconstruction \cite{liao2024fishspeech}.

pyttsx3 is an offline Python text-to-speech library that serves as an interface to the TTS engines installed in the operating system. Speech synthesis is performed by selecting a supported operating-system backend (e.g., eSpeak/eSpeak-NG on Linux, SAPI5 on Windows, or NSSpeechSynthesizer/AVSpeechSynthesizer on macOS) and calling the library API to generate speech output.

RHVoice is a free and open-source speech synthesizer based on statistical parametric synthesis and provides Russian and 10 additional supported languages. \cite{rhvoice_github}
It is available on Windows, GNU/Linux, and Android, and is compatible with standard platform TTS interfaces, including SAPI5 (Windows), Speech Dispatcher (GNU/Linux), and Android text-to-speech APIs. \cite{rhvoice_github}

Silero TTS~\cite{silero2021} is an end-to-end neural TTS system supporting multiple languages and speakers.
For Russian, it provides automatic stress placement and homograph handling, and includes five speakers (aidar, baya, kseniya, xenia, eugene).

Fairseq Transformer TTS is a Russian TTS checkpoint from the fairseq $S^2$ suite~\cite{li2019neural}.
It was pre-trained on Common Voice v7 and fine-tuned on CSS10, providing a single male speaker.
The system follows a two-stage synthesis pipeline: a Transformer-based acoustic model predicts intermediate acoustic features from the input text, which are subsequently converted into waveform by a neural vocoder.

A Russian fine-tuned SpeechT5 TTS checkpoint trained on Common Voice 13 is considered in this study~\cite{ao2022speecht5}.
The model expects Russian transliteration as input and employs an encoder-decoder Transformer to predict an intermediate acoustic representation, followed by waveform synthesis using a HiFi-GAN vocoder.

Vosk-TTS provides five selectable speakers (three female and two male).
Text is converted into phoneme sequences; an ODE-based decoder trained with conditional flow matching predicts acoustic features, followed by vocoder-based waveform reconstruction.

Edge TTS provides Microsoft Edge neural voices and follows a three-stage pipeline comprising text normalization and phonemization, a neural acoustic model that predicts prosodic and spectral features (e.g., intonation and stress), and a neural vocoder that synthesizes the waveform. VK Cloud Voice, SaluteSpeech, and ElevenLabs TTS expose cloud-based neural TTS via APIs with configurable synthesis parameters; VK Cloud Voice supports speech-rate control and returns PCM/MP3/Opus audio with six speakers, SaluteSpeech supports prosody control via API parameters and SSML with six Russian and one English voice, and ElevenLabs TTS provides controls for sampling variability, voice-similarity preservation, and style conditioning. For Edge TTS, ru-RU-DmitryNeural and ru-RU-SvetlanaNeural were used.

\subsection{Spoof generation protocol}

\Figure[t!](topskip=0pt, botskip=0pt, midskip=0pt)[width=\textwidth]{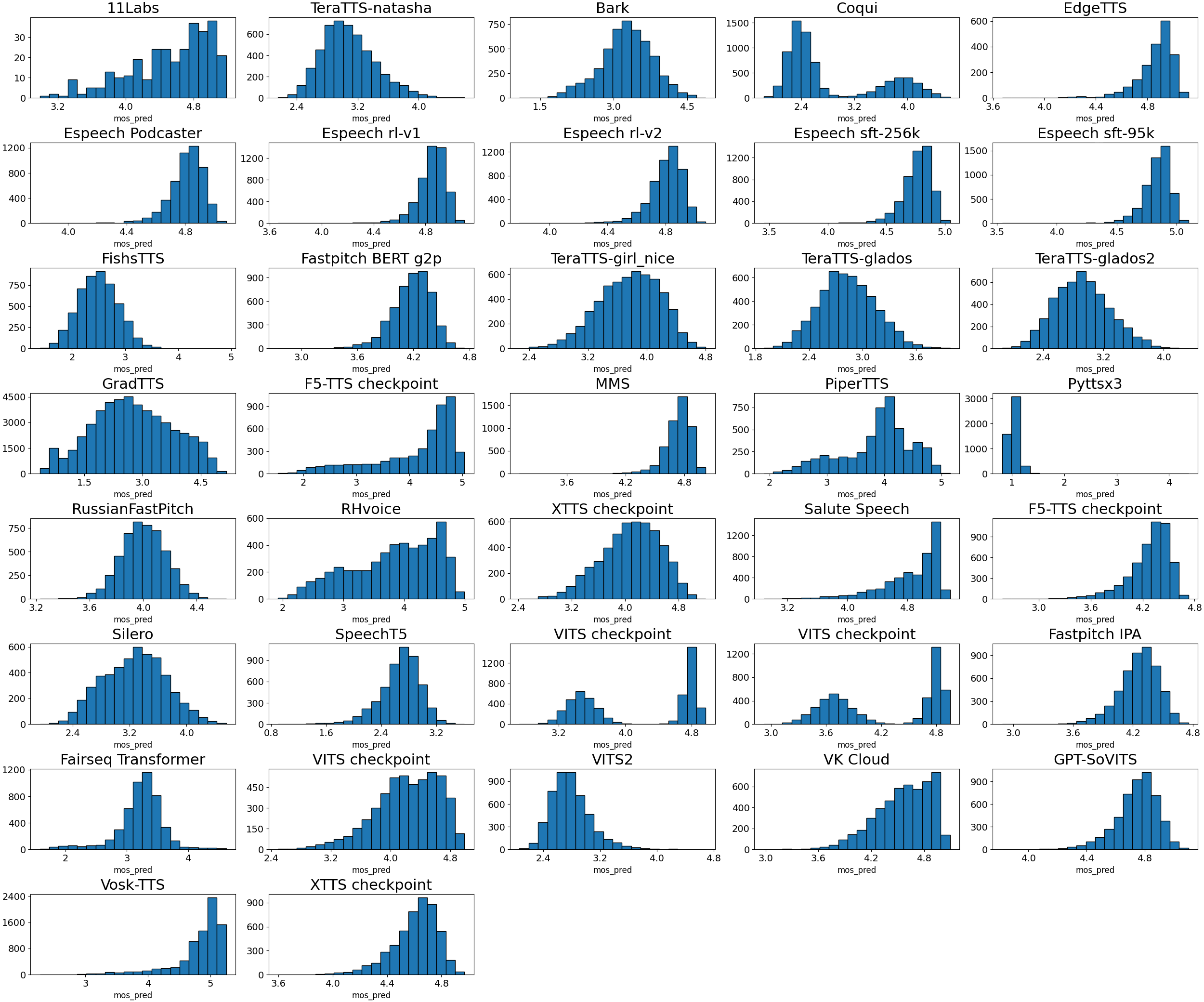}
{Predicted MOS (NISQA) distributions for spoof speech across TTS systems.\label{fig:spoof_mos}}

\Figure[t!](topskip=0pt, botskip=0pt, midskip=0pt)[width=\textwidth]{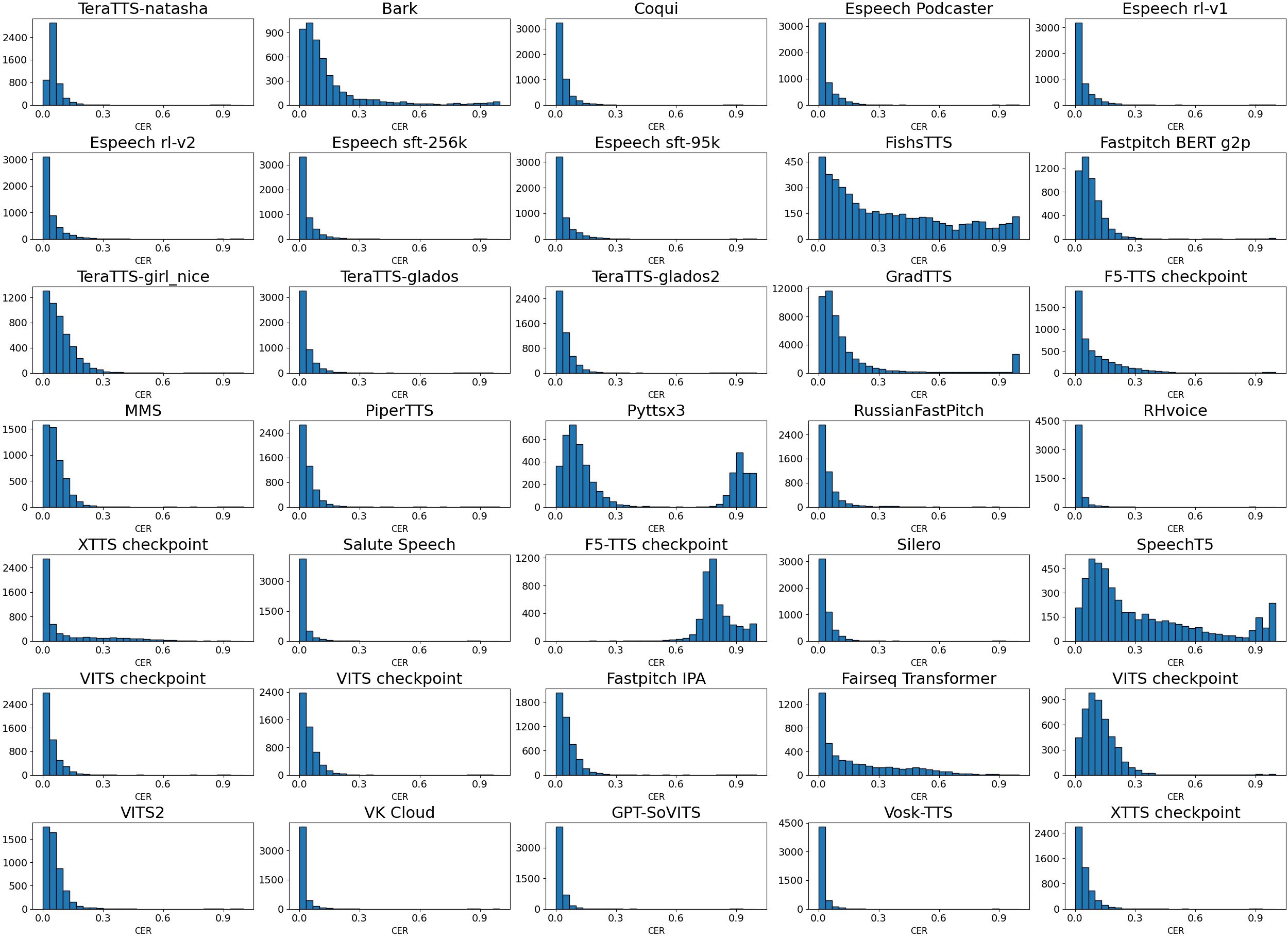}
{ASR-based character error rate (CER) distributions for spoof speech across TTS systems.\label{fig:spoof_cer}}

\label{sec:spoof_protocol}
For the spoof subset of \textbf{RuASD}, spoof speech was synthesized using 37 Russian-capable TTS systems.
To balance the contribution of different generators, we targeted approximately 5{,}000 utterances per system, thereby limiting dataset-size bias and preventing the corpus from being dominated by a small number of large-scale generators.
The realized number of recordings and total duration for each system are reported in Table~\ref{tab:spoof_model_stats}.

Input texts were sampled from the Russian side of the United Nations Parallel Corpus (UNPC) distributed via OPUS\footnote{\url{https://opus.nlpl.eu/datasets/UNPC}}.

\begin{table*}[t]
\centering
\caption{RuASD spoof subset: per-TTS statistics and objective quality descriptors.}
\label{tab:spoof_model_stats}
\setlength{\tabcolsep}{1.5pt}
\renewcommand{\arraystretch}{1.1}

\begin{adjustbox}{width=\textwidth}

\begin{tabular}{l c c c c c c c c c c}

\toprule
\textbf{TTS system} & \textbf{CER} & 
\textbf{MOS} & \textbf{Dis} & 
\textbf{Col} & \textbf{Loud} & \textbf{Noi} & 
\textbf{Total records} & \textbf{Total duration (h)} & 
\textbf{Average duration (s)} & Link \\
\midrule

\textit{Espeech Podcaster} & \uline{0.04$\pm$0.003} & 4.81$\pm$0.003 & 4.69$\pm$0.003 & 4.48$\pm$0.003 & 4.37$\pm$0.005 & 4.50$\pm$0.002 & 5000 & 14.12 & 10.17 & \href{https://hf.co/ESpeech/ESpeech-TTS-1_podcaster}{link} \\
\textit{Espeech RL-V1} & \uline{0.04$\pm$0.003} & \uline{4.85$\pm$0.003} & \textbf{4.72$\pm$0.003} & \uline{4.52$\pm$0.003} & 4.47$\pm$0.004 & \uline{4.53$\pm$0.002} & 5000 & 14.12 & 10.17 & \href{https://hf.co/ESpeech/ESpeech-TTS-1_RL-V1}{link} \\
\textit{Espeech RL-V2} & 0.05$\pm$0.005 & 4.80$\pm$0.003 & 4.68$\pm$0.002 & 4.47$\pm$0.003 & 4.42$\pm$0.005 & 4.50$\pm$0.002 & 5000 & 14.12 & 10.17 & \href{https://hf.co/ESpeech/ESpeech-TTS-1_RL-V1}{link}\\
\textit{Espeech SFT-95k} & \uline{0.04$\pm$0.003} & 4.83$\pm$0.003 & \uline{4.71$\pm$0.003} & 4.50$\pm$0.003 & 4.48$\pm$0.004 & \textbf{4.54$\pm$0.002} & 5000 & 14.12 & 10.17 & \href{https://hf.co/ESpeech/ESpeech-TTS-1_SFT-95K}{link}\\
\textit{Espeech SFT-256k} & \uline{0.04$\pm$0.003} & 4.76$\pm$0.004 & 4.67$\pm$0.003 & 4.44$\pm$0.003 & 4.41$\pm$0.005 & \uline{4.53$\pm$0.002} & 5000 & 14.12 & 10.17 & \href{https://hf.co/ESpeech/ESpeech-TTS-1_SFT-95K}{link}\\
\textit{F5-TTS checkpoint} & 0.13$\pm$0.01 & 4.11$\pm$0.02 & 4.22$\pm$0.01 & 4.06$\pm$0.01 & 3.98$\pm$0.01 & 3.78$\pm$0.02 & 5000 & 12.80 & 9.22 & \href{https://hf.co/Misha24-10/F5-TTS_RUSSIAN}{link}\\
\textit{F5-TTS checkpoint} & 0.85$\pm$0.007 & 4.27$\pm$0.007 & 4.44$\pm$0.004 & 4.27$\pm$0.004 & 4.24$\pm$0.004 & 4.01$\pm$0.005 & 5009 & \uline{44.36} & 31.88 & \href{https://hf.co/hotstone228/F5-TTS-Russian}{link}\\
\textit{VITS checkpoint} & 0.13$\pm$0.006 & 4.22$\pm$0.01 & 4.21$\pm$0.01 & 3.81$\pm$0.01 & 4.26$\pm$0.007 & 4.23$\pm$0.006 & 5000 & 13.43 & 9.67 & \href{https://hf.co/joefox/tts_vits_ru_hf}{link}\\
\textit{PiperTTS} & 0.05$\pm$0.005 & 3.87$\pm$0.02 & 4.42$\pm$0.006 & 4.02$\pm$0.007 & 4.04$\pm$0.01 & 4.05$\pm$0.006 & 5045 & 14.02 & 10.01 & \href{https://github.com/rhasspy/piper}{link}\\
\textit{TeraTTS-natasha} & 0.06$\pm$0.002 & 3.09$\pm$0.01 & 3.80$\pm$0.006 & 3.58$\pm$0.005 & 3.62$\pm$0.006 & 3.59$\pm$0.006 & 5000 & 14.22 & 10.24 & \href{https://hf.co/TeraTTS/natasha-g2p-vits}{link}\\
\textit{TeraTTS-girl\_nice} & 0.10$\pm$0.01 & 3.76$\pm$0.01 & 4.01$\pm$0.01 & 3.82$\pm$0.01 & 4.11$\pm$0.006 & 4.09$\pm$0.006 & 5500 & 19.15 & 12.54 & \href{https://hf.co/TeraTTS/girl_nice-g2p-vits}{link}\\
\textit{TeraTTS-glados} & \uline{0.04$\pm$0.003} & 2.83$\pm$0.01 & 3.82$\pm$0.006 & 3.28$\pm$0.004 & 3.22$\pm$0.005 & 3.71$\pm$0.003 & 5015 & 14.46 & 10.38 & \href{https://hf.co/TeraTTS/glados-g2p-vits}{link}\\
\textit{TeraTTS-glados2} & 0.05$\pm$0.006 & 2.92$\pm$0.01 & 3.94$\pm$0.006 & 3.46$\pm$0.007 & 2.87$\pm$0.006 & 3.69$\pm$0.004 & 5000 & 13.55 & 9.75 & \href{https://hf.co/TeraTTS/glados2-g2p-vits}{link}\\
\textit{MMS} & 0.07$\pm$0.004 & 4.73$\pm$0.004 & 4.71$\pm$0.004 & 4.33$\pm$0.004 & \uline{4.56$\pm$0.003} & 4.44$\pm$0.002 & 5000 & 15.49 & 11.16 & \href{https://hf.co/facebook/mms-tts-rus}{link}\\
\textit{VITS checkpoint} & 0.05$\pm$0.002 & 4.23$\pm$0.02 & 4.39$\pm$0.01 & 4.03$\pm$0.01 & 4.23$\pm$0.01 & 4.02$\pm$0.01 & 5000 & 14.82 & 10.67 & \href{https://hf.co/utrobinmv/tts_ru_free_hf_vits_low_multispeaker}{link}\\
\textit{VITS checkpoint} & \uline{0.04$\pm$0.001} & 4.13$\pm$0.02 & 4.35$\pm$0.01 & 4.01$\pm$0.01 & 4.16$\pm$0.01 & 4.03$\pm$0.01 & 5000 & 15.85 & 11.41 & \href{https://hf.co/utrobinmv/tts_ru_free_hf_vits_high_multispeaker}{link}\\
\textit{VITS2 checkpoint} & 0.06$\pm$0.002 & 2.79$\pm$0.01 & 3.79$\pm$0.004 & 3.36$\pm$0.004 & 3.30$\pm$0.005 & 3.62$\pm$0.005 & 5013 & 12.56 & 9.02 & \href{https://hf.co/frappuccino/vits2_ru_natasha}{link}\\
\textit{GPT-SoVITS checkpoint} & \textbf{0.02$\pm$0.001} & 4.74$\pm$0.004 & 4.50$\pm$0.003 & 4.35$\pm$ 0.003 & 4.54$\pm$0.002 & 4.46$\pm$0.004 & 5000 & 13.49 & 9.71 & \href{https://hf.co/alphacep/vosk-tts-ru-gpt-sovits}{link}\\
\textit{CoquiTTS} & \uline{0.04$\pm$0.003} & 2.85$\pm$0.02 & 4.06$\pm$0.006 & 3.48$\pm$0.007 & 3.21$\pm$0.01 & 2.98$\pm$0.02 & 7291 & 23.12 & 11.42 & \href{https://hf.co/coqui/XTTS-v2}{link}\\
\textit{XTTS checkpoint} & 0.11$\pm$0.005 & 4.08$\pm$0.01 & 4.24$\pm$0.006 & 4.11$\pm$0.005 & 4.37$\pm$0.004 & 3.84$\pm$0.007 & 5017 & 14.24 & 10.22 & \href{https://hf.co/NeuroDonu/RU-XTTS-DonuModel}{link}\\
\textit{XTTS checkpoint} & 0.05$\pm$0.001 & 4.60$\pm$0.005 & 4.65$\pm$0.002 & 4.39$\pm$0.002 & 4.41$\pm$0.004 & 4.37$\pm$0.003 & 5000 & 14.15 & 10.19 & \href{https://hf.co/omogr/xtts-ru-ipa}{link}\\
\textit{Fastpitch IPA} & 0.09$\pm$0.006 & 4.18$\pm$0.006 & 4.36$\pm$0.005 & 3.85$\pm$0.005 & 4.25$\pm$0.004 & 4.11$\pm$0.005 & 5000 & 13.09 & 9.42 & \href{https://hf.co/bene-ges/tts_ru_ipa_fastpitch_ruslan}{link}\\
\textit{Fastpitch BERT g2p} & 0.06$\pm$0.005 & 4.26$\pm$0.006 & 4.41$\pm$0.004 & 3.93$\pm$0.005 & 4.29$\pm$0.004 & 4.03$\pm$0.003 & 4954 & 13.44 & 9.77 & \href{https://hf.co/bene-ges/ru_g2p_ipa_bert_large}{link}\\
\textit{RussianFastPitch} & 0.05$\pm$0.002 & 4.00$\pm$0.005 & 4.40$\pm$0.003 & 3.91$\pm$0.004 & 4.19$\pm$0.004 & 3.85$\pm$0.007 & 5000 & 15.88 & 11.43 & \href{https://github.com/safonovanastya/RussianFastPitch}{link}\\
\textit{Bark} & 0.18$\pm$0.01 & 3.26$\pm$0.01 & 4.11$\pm$0.01 & 3.59$\pm$0.01 & 3.75$\pm$0.01 & 3.23$\pm$0.03 & 5000 & 17.00 & 12.21 & \href{https://hf.co/suno/bark-small}{link} \\
\textit{GradTTS} & 0.16$\pm$0.002 & 2.75$\pm$0.01 & 3.52$\pm$0.007 & 3.16$\pm$0.007 & 3.5$\pm$0.007 & 3.31$\pm$0.01 &\textbf{50108} & \textbf{124.33} & 8.93 & \href{https://github.com/huawei-noah/Speech-Backbones/tree/main/Grad-TTS}{link} \\
\textit{FishTTS} & 0.61$\pm$0.04 & 2.50$\pm$0.01 & 3.35$\pm$0.01 & 2.83$\pm$0.007 & 3.36$\pm$0.007 & 2.64$\pm$0.01 & 5015 & 11.70 & 8.36 & \href{https://hf.co/fishaudio/fish-speech-1.5}{link}\\
\textit{Pyttsx3} & 0.48$\pm$0.02 & 1.04$\pm$0.004 & 1.71$\pm$0.005 & 2.54$\pm$0.004 & 3.29$\pm$0.005 & 3.80$\pm$0.005 & 5005 & 14.66 & 10.54 & \href{https://github.com/nateshmbhat/pyttsx3}{link}\\
\textit{RHVoice} & \textbf{0.02$\pm$0.001} & 3.82$\pm$0.02 & 4.11$\pm$0.01 & 3.75$\pm$0.01 & 3.72$\pm$0.01 & 3.99$\pm$0.007 & 5106 & 17.07 & 12.03 & \href{https://github.com/RHVoice/RHVoice}{link}\\
\textit{Silero} & \uline{0.04$\pm$0.002} & 3.27$\pm$0.01 & 3.91$\pm$0.007 & 3.31$\pm$0.007 & 3.50$\pm$0.01 & 3.66$\pm$0.01 & 4979 & 13.17 & 9.52 & \href{https://github.com/snakers4/silero-models}{link}\\
\textit{Fairseq Transformer} & 0.45$\pm$0.03 & 3.22$\pm$0.01 & 4.12$\pm$0.01 & 3.14$\pm$0.01 & 3.44$\pm$0.01 & 3.26$\pm$0.01 & 5000 & 11.92 & 8.58 & \href{https://hf.co/facebook/tts_transformer-ru-cv7_css10}{link}\\
\textit{SpeechT5} & 0.38$\pm$0.02 & 2.67$\pm$0.01 & 3.95$\pm$0.01 & 2.80$\pm$0.01 & 3.45$\pm$0.007 & 2.68$\pm$0.01 & 5063 & 18.21 & 12.95 & \href{https://hf.co/voxxer/speecht5_finetuned_commonvoice_ru_translit}{link}\\
\textit{Vosk-TTS} & \textbf{0.02$\pm$0.001} & 4.81$\pm$0.01 & 4.65$\pm$0.006 & 4.41$\pm$0.007 & 4.47$\pm$0.008 & 4.37$\pm$0.007 & \uline{7844} & 37.54 & 17.23 & \href{https://github.com/alphacep/vosk-tts}{link}\\
\textit{EdgeTTS} & -- & \textbf{4.86$\pm$0.007} & 4.65$\pm$0.003 & \textbf{4.53$\pm$0.003} & \textbf{4.66$\pm$0.002} & 4.50$\pm$0.002 & 2024 & 13.53 & 24.07 & \href{https://github.com/rany2/edge-tts}{link} \\
\textit{VK Cloud} & \textbf{0.02$\pm$0.001} & 4.56$\pm$0.01 & 4.52$\pm$0.007 & 4.34$\pm$0.006 & 4.20$\pm$0.01 & 4.13$\pm$0.006 & 4975 & 14.66 & 10.61 & \href{https://cloud.vk.com/}{link} \\
\textit{SaluteSpeech} & \textbf{0.02$\pm$0.001} & \textbf{4.86$\pm$0.01} & 4.45$\pm$0.01 & 4.31$\pm$0.01 & 4.44$\pm$0.01 & 4.37$\pm$0.006 & 4997 & 13.95 & 10.05 & \href{https://developers.sber.ru/portal/products/smartspeech}{link} \\
\textit{ElevenLabs} & -- & 4.52$\pm$0.05 & 4.48$\pm$0.03 & 4.24$\pm$0.04 & 4.25$\pm$0.05 & 4.25$\pm$0.05 & 306 & 1.22 & 14.30 & \href{https://elevenlabs.io/}{link} \\
\bottomrule
\end{tabular}

\end{adjustbox}

\begin{tablenotes}
\footnotesize
\item \textit{Note:} CER is reported only when a reference transcript is available; NISQA provides predicted MOS and the quality dimensions Discontinuity (Dis), Coloration (Col), and Loudness \cite{mittag2021nisqa}.
\end{tablenotes}

\end{table*}

\paragraph{Data storage format.}
Synthesized utterances were stored in the original audio format and sampling rate produced by each TTS system, without additional post-processing.
This choice preserves the raw outputs of each generator and keeps the dataset closer to practical attack pipelines, where the defender may receive audio produced and exported in heterogeneous formats.
At the same time, model evaluation requires a unified input format (Sec.~\ref{sec:experimental_setup}); therefore, all recordings are resampled prior to inference, which may interact with generator-specific artifacts and is discussed as a limitation.

\paragraph{Objective spoof characterization.}
For every synthesized utterance, we computed (i) an ASR-based character error rate (CER) using Whisper-medium~\cite{radford2023whisper} and (ii) the predicted MOS using NISQA~\cite{mittag2021nisqa}. For a subset of spoof recordings, an exact normalized reference text is not available in a form suitable for CER computation (e.g., text normalization/SSML or missing per-utterance input logs), and CER is therefore omitted.
Fig.~\ref{fig:spoof_mos} and Fig.~\ref{fig:spoof_cer} illustrate cross-generator variability in both MOS and CER.
We report MOS and CER as objective dataset descriptors (sanity checks) rather than as proxies for spoof ``strength'' or detector difficulty.

\subsection{Bona fide sources}
\label{sec:bonafide_sources}

\begin{figure*}[t]
\centering

\begin{minipage}{\textwidth}
    \centering
    \includegraphics[width=0.75\textwidth]{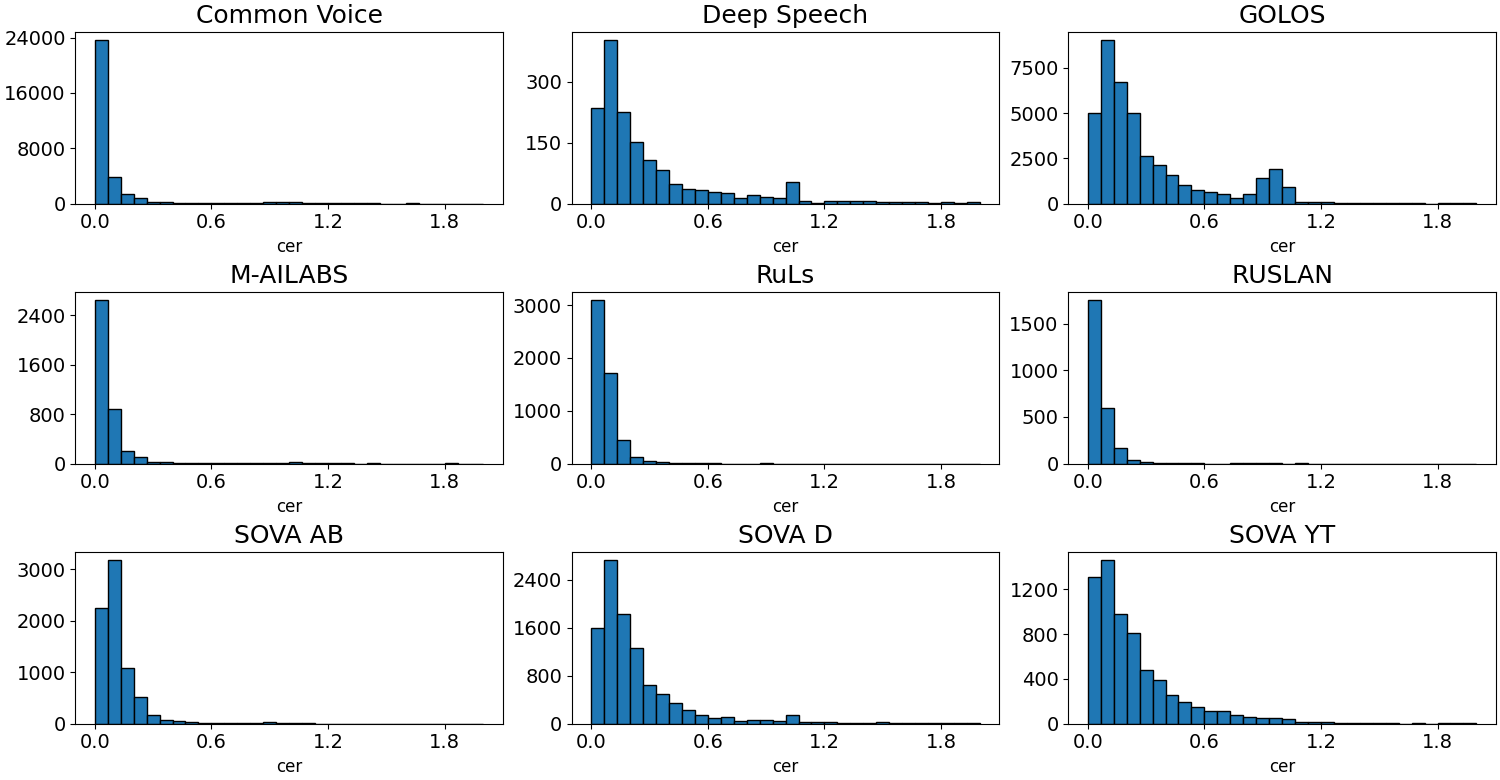}
    
    \vspace{2mm}
    {\footnotesize (a) CER.}
\end{minipage}

\vspace{4mm}

\begin{minipage}{\textwidth}
    \centering
    \includegraphics[width=\textwidth]{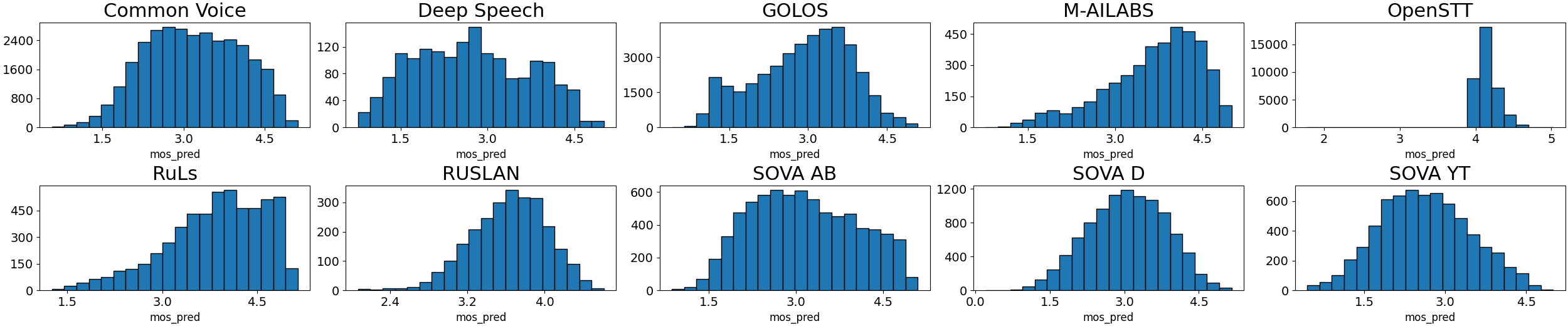}
    
    \vspace{2mm}
    {\footnotesize (b) MOS (NISQA).}
\end{minipage}

\caption{Objective descriptors for the bona fide subset across source corpora.}
\label{fig:bonafide_descriptors}
\end{figure*}

As sources of bona fide Russian speech, we use Deep Speech~\cite{fedoseev2017deepspeech}, GOLOS~\cite{karpov2021golos}, M-AILABS~\cite{MAILABS_2017}, OpenSTT~\cite{bondarenko2019openstt,borodin2025datacentricframeworkaddressingphonetic}, RuLS~\cite{ruls2020}, RUSLAN~\cite{ruslan2020}, Mozilla Common Voice~\cite{ardila2020common}, and SOVA~\cite{chadeeva2024sova}.
To better match deployment conditions, we intentionally pool heterogeneous bona fide audio (read speech, crowdsourced and far-field recordings, and in-the-wild sources), which increases variability in channel characteristics and transcript quality and avoids an overly curated bona fide domain.

Deep Speech was automatically collected from YouTube videos with subtitles by extracting audio and subtitle text, segmenting speech using WebRTC VAD, and aligning transcripts via timestamps.We use a 3-hour random subsample from the original 6K-hour corpus.

GOLOS is a manually annotated Russian speech corpus covering both crowdsourced and far-field speech ($\sim$1,240 h), collected via Yandex.Toloka and SberPortal sessions with simulated far-field conditions. In this work, we use a 72 h subset.

The M‑AILABS Speech Dataset is based primarily on LibriVox audiobooks and texts from Project Gutenberg. The Russian subset contains 46 h 47 min of audio. In this work, we used a random subsample with a total duration of 9.34 hours.

OpenSTT is a large-scale public Russian ASR corpus aggregated from multiple domains (e.g., radio, public talks, audiobooks, YouTube, and telephone calls), comprising 20{,}108 hours. In this work, we used a random subsample totaling 56.98 hours.

Russian LibriSpeech (RuLS) is an open corpus of Russian read speech based on publicly available LibriVox audiobooks. Its total duration is approximately 98 hours. In our work, we used a random subsample with a total duration of 10.14 hours.

RUSLAN is an open Russian single-speaker TTS corpus containing 22{,}200 text-audio pairs (30+ hours), recorded in a sound-isolated room with noise-reduction hardware. In this work, we used 3.68 hours of audio.

Mozilla Common Voice is a crowdsourced multilingual corpus of transcribed speech, with community-based collection and validation. In this work, we used the Russian Spontaneous Speech 2.0 subset (426 clips, 3 h total, 2 h validated) from 12 speakers.

The SOVA dataset is an open multilingual speech corpus with approximately 32{,}328 hours of audio.
From its Russian partition, which covers RuYoutube (YouTube recordings, 17{,}451 hours), RuAudiobooksDevices (audiobooks/reading, 298 hours), and RuDevices (device recordings, 101 hours), we selected 30.44 hours for bona fide speech.

\subsection{Bona fide data selection protocol}
\label{sec:bonafide_protocol}
To construct the bona fide subset, we collected candidate utterances from multiple Russian speech corpora and then selected a controlled subset by constraining the total amount of selected audio to 10 GB.

\paragraph{Candidate pool and selection.}
We merged recordings from all selected sources into a single candidate pool and extracted lightweight metadata for each utterance, including the source corpus identifier, split label, speaker identifier, utterance duration, transcript and file size in bytes.
These metadata were used to (i) enforce per-dataset size constraints under a fixed 10~GB total limit, (ii) balance the contribution of different splits and speakers, (iii) control the distribution of utterance durations, and (iv) bound the total size of the selected subset via the cumulative file size.

For datasets exposing multiple splits, we enforced an approximately uniform per-split size constraint and performed selection independently within each split.
Whenever a split could not satisfy its constraint (e.g., due to limited availability or rounding), the shortfall was compensated using unused candidates from the remaining splits of the same dataset.
Selection within each split was deterministic under a fixed random seed, ensuring that repeated runs on the same inputs yield the same selected subset.
To increase diversity, candidates were stratified by speaker and, when duration metadata was available, by duration bins: \texttt{0--3s}, \texttt{3--10s}, \texttt{10--30s}, \texttt{30--120s}, \texttt{120s+}.
Selection was then performed via round-robin sampling over groups defined by the key \texttt{(split, speaker, duration bucket)}: at each step, one item was taken from the next group until the cumulative size of the selected files reached the target size for the corresponding split.

\paragraph{Objective bona fide characterization.}
\begin{table*}[t]
\centering
\caption{RuASD bona fide subset: per-corpus statistics and objective quality descriptors.}
\label{tab:bonafide_dataset_stats}
\setlength{\tabcolsep}{2.5pt}

\begin{adjustbox}{width=\textwidth}

\begin{tabular}{l c c c c c c c c c}
\toprule
\textbf{Speech Datasets} & \textbf{CER} & 
\textbf{MOS} & \textbf{Dis} & 
\textbf{Col} & \textbf{Loud} & \textbf{Noi} &
\textbf{Total records} & \textbf{Total duration (h)} & 
\textbf{Average duration (s)} \\
\midrule
\textit{Deep Speech} & 1.04$\pm$0.56 & 2.73$\pm$0.05 & 3.65$\pm$0.03 & 3.02$\pm$0.03 & 3.33$\pm$0.04 & 2.64$\pm$0.05 & 1663 & 3.01 & 6.52 \\
\textit{GOLOS} & 0.32$\pm$0.005 & 2.91$\pm$0.01 & 3.84$\pm$0.005 & 3.19$\pm$0.006 & 3.15$\pm$0.01 & 2.95$\pm$0.007 & \textbf{40648} & \textbf{72.10} & 6.39 \\
\textit{M-AILABS} & 0.09$\pm$0.008 & 3.66$\pm$0.025 & 3.49$\pm$0.026 & 3.85$\pm$0.01 & 3.91$\pm$0.02 & \textbf{4.14$\pm$0.01} & 3996 & 9.34 & 8.42 \\
\textit{RuLS} & \uline{0.08$\pm$0.005} & \uline{3.85$\pm$0.02} & 4.11$\pm$0.02 & 3.72$\pm$0.02 & 4.11$\pm$0.01 & \uline{3.90$\pm$0.02} & 5505 & 10.14 & 6.63 \\
\textit{RUSLAN} & \textbf{0.06$\pm$0.003} & 3.63$\pm$0.02 & \uline{4.39$\pm$0.01} & \uline{3.94$\pm$0.01} & \uline{4.23$\pm$0.01} & 3.77$\pm$0.02 & 2600 & 3.68 & 5.1 \\
\textit{SOVA-AB} & 0.13$\pm$0.01 & 3.15$\pm$0.02 & 3.72$\pm$0.02 & 3.25$\pm$0.02 & 3.55$\pm$0.015 & 3.21$\pm$0.02 & 7457 & 10.14 & 4.9 \\
\textit{SOVA-D} & 0.38$\pm$0.08 & 3.05$\pm$0.015 & 3.86$\pm$0.01 & 3.15$\pm$0.01 & 3.45$\pm$0.01 & 2.87$\pm$0.01 & 10082 & 10.14 & 3.62 \\
\textit{SOVA-YT} & 0.32$\pm$0.03 & 2.62$\pm$0.02 & 3.49$\pm$0.02 & 2.86$\pm$0.02 & 3.16$\pm$0.02 & 2.86$\pm$0.02 & 6648 & 10.16 & 5.5 \\
\textit{Mozilla Common Voice} & 0.09$\pm$0.009 & 3.18$\pm$0.01 & 3.79$\pm$0.008 & 3.45$\pm$0.007 & 3.73$\pm$0.007 & 3.45$\pm$0.008 & 31456 & 48.38 & 5.54 \\
\textit{OpenSTT} & -- & \textbf{4.15$\pm$0.001} & \textbf{4.42$\pm$0.002} & \textbf{4.04$\pm$0.002} & \textbf{4.25$\pm$0.002} & 3.61$\pm$0.005 & \uline{37042} & \uline{56.98} & 5.54 \\
\bottomrule
\end{tabular}

\end{adjustbox}

\begin{tablenotes}
\footnotesize
\item \textit{Note:} CER is reported only when a reference transcript is available; NISQA provides predicted MOS and the quality dimensions Discontinuity (Dis), Coloration (Col), and Loudness \cite{mittag2021nisqa}.
\end{tablenotes}
\end{table*}

In addition to summary statistics (Table~\ref{tab:bonafide_dataset_stats}), we report CER and NISQA-predicted MOS distributions per source corpus (Fig.~\ref{fig:bonafide_descriptors}(a) and Fig.~\ref{fig:bonafide_descriptors}(b)).

\paragraph{Comparison of bona fide and spoof subsets.}
As a result, we obtained a bona fide subset with a controlled contribution from each source dataset, an approximately balanced distribution across splits (when provided by the sources), and increased speaker and duration diversity whenever such metadata was available.
Comparing the objective descriptors across the bona fide (Table~\ref{tab:bonafide_dataset_stats}) and spoof (Table~\ref{tab:spoof_model_stats}) subsets reveals distinct quality profiles.
Bona fide sources span a broad range of perceptual quality (Table~\ref{tab:bonafide_dataset_stats}), ranging from mid-range MOS values in several corpora to higher MOS in more curated read-speech datasets and in OpenSTT, which exhibits a high MOS but does not provide reference transcripts for CER computation.
ASR-based intelligibility for bona fide speech is also heterogeneous among corpora with available references, with CER ranging from 0.06 up to 1.04 (Table~\ref{tab:bonafide_dataset_stats}).

In contrast, spoof speech shows strong generator dependence (Table~\ref{tab:spoof_model_stats} and Fig.~\ref{fig:spoof_mos}): several modern neural and cloud TTS systems yield narrow, high-MOS distributions near the upper end of the scale, while other generators remain in the mid- to low-MOS range (e.g., classical/offline pipelines and some open-source models).
The CER distributions for spoof speech (Fig.~\ref{fig:spoof_cer}) are typically concentrated near low error rates for many generators, yet the spread and occasional heavier tails highlight that intelligibility is not uniformly high across all synthesis pipelines (Table~\ref{tab:spoof_model_stats}).
Overall, the joint analysis of summary statistics (Tables~\ref{tab:spoof_model_stats}--\ref{tab:bonafide_dataset_stats}) and distributions (Figs.~\ref{fig:bonafide_descriptors}(a)--\ref{fig:bonafide_descriptors}(b)) indicates that the dataset includes both highly natural and clearly degraded spoofed speech, while the bona fide portion reflects diverse real-world recording conditions and annotation quality rather than a single homogeneous capture setup.
CER is reported only when reference transcripts are available and is omitted otherwise, consistent with the per-source and per-generator reporting (Tables~\ref{tab:spoof_model_stats}--\ref{tab:bonafide_dataset_stats}).

\subsection{Data Augmentation}
To assess the robustness of the models against data perturbations and realistic channel distortions, three types of augmentation were used: adding reverberation by convolving the signal with room impulse responses (RIRs), adding additive noise with a specified SNR range (using MUSAN dataset), and simulating communication-channel distortions by re-encoding the signal with various speech codecs (mp3, opus\_8khz, opus\_16k, g722, alaw, mulaw, speex\_8khz, amr) and then resampling the resulting signal to 16kHz. These augmentations emulate common real-world dissemination pipelines for voice traffic and user-generated audio, where recordings are affected by room acoustics, background interference, and platform- or codec-driven transcoding.
Accordingly, we treat the augmentation conditions as controlled simulations of deployment shifts rather than as generic data augmentation for training.

Codec augmentations were implemented by sequentially encoding the audio signal into a compressed format and then decoding it back to PCM (WAV), which simulates transmission/transcoding artifacts in real communication channels and platforms. Encoding and decoding were performed using pydub (ffmpeg backend). For MP3/Opus/Speex/AMR, the bitrate was randomized within predefined ranges, whereas fixed encoding parameters were used for the G.722 and G.711 a-law/mu-law codecs (G.722 was encoded at a fixed 64kbit/s bitrate, and G.711 a-law/mu-law was applied in the 8kHz branch without bitrate randomization). The sampling rate was also forced before encoding (8kHz for narrowband scenarios and 16kHz for wideband), after which the decoded signal was resampled to 16kHz to unify model inputs and saved as WAV (PCM16).

Source recordings for augmentation were selected from an aggregated pool of audio files comprising both bona fide recordings (collected as detailed in Sec.~\ref{sec:bonafide_protocol}) and TTS-synthesized speech (generated according to the methodology outlined inSec.~\ref{sec:spoof_protocol}). A distinct subset of files was formed from this shared pool for each specific augmentation combination. For each augmentation condition listed in Table~\ref{tab:models_aug}, we generated 3{,}000 augmented utterances by deterministically sampling source recordings with a fixed random seed. Specifically, we aimed to approximately match the label distribution (bona fide vs.spoof) observed in the full dataset and, for spoof samples, to preserve the relative contribution of different TTS generators. For codec-based conditions, we assigned a codec to each utterance via an explicit per-codec count specification. The size of the augmented subset generated for a given condition, \(N\), was defined in the configuration either (i) by a fixed \texttt{count} (non-codec chains) or (ii) as \(N=\sum_c n_c\) via \texttt{codec\_counts} (codec-based chains), where \(n_c\) is the requested number of samples for codec \(c\). Finally, we formed a codec sequence of length \(N\) by repeating each codec label \(c\) exactly \(n_c\) times, shuffled this sequence using the same seeded RNG, and applied the assigned codec to each utterance at the codec step of the augmentation chain.

\section{Experimental Setup}
\label{sec:experimental_setup}
This section defines a reproducible evaluation protocol for \textbf{RuASD} intended to isolate robustness effects under realistic channel and post-processing shifts.
In addition to clean evaluation, we emphasize robustness-oriented assessment: the same set of detectors is evaluated under controlled degradations (noise, reverberation, and transcoding) while keeping pre-processing and score computation consistent across models whenever possible.

\subsection{Evaluation Metrics}
To objectively assess the quality of the synthesized spoofing attacks and the performance of detection models on \textbf{RuASD}, we employ two sets of metrics: one for speech quality assessment and one for countermeasure performance.
\subsubsection{Speech Quality and Artifact Assessment}
We evaluate the perceptual quality and intelligibility of the generated spoofing attacks to ensure they represent non-trivial threats.
\begin{itemize}
    \item \textbf{NISQA (Non-Intrusive Speech Quality Assessment):} We use the NISQA model~\cite{mittag2021nisqa} to predict the Mean Opinion Score (MOS) and specific quality dimensions (Noisiness, Coloration, Discontinuity, and Loudness) directly from waveforms, without requiring reference clean recordings. This verifies that the generated attacks maintain high perceptual quality and are not trivially degraded.
    \item \textbf{ASR-based CER (Character Error Rate):} To assess intelligibility and phonetic fidelity, we transcribe all bona fide and spoofed utterances using the \texttt{whisper-medium} model~\cite{radford2023whisper}. The Character Error Rate (CER) is calculated between the recognized text and the ground truth transcriptions. Lower CER indicates higher intelligibility, but CER is only computed when reference transcripts are available, and should be interpreted as an auxiliary descriptor rather than a primary corpus objective.
\end{itemize}

\subsubsection{Anti-Spoofing Performance}
Detection performance is evaluated using standard binary classification metrics. We report Accuracy (Acc), Precision (Pr), Recall (Rec), F1-score, and the Area Under the ROC Curve (ROC-AUC). Additionally, we report the Equal Error Rate (EER), a standard biometric metric defined as the error rate at the operating point where the false acceptance rate (FAR) equals the false rejection rate (FRR). EER is reported as a standard biometric operating point that is less sensitive to a particular decision threshold choice.

\subsection{Benchmarked Countermeasures}
We benchmark three groups of anti-spoofing countermeasures: (i) \textit{Convolutional and temporal modeling backbones} (Res2TCNGuard, ResCapsGuard~\cite{borodin2024_rescaps_res2tcn}, Nes2Net~\cite{liu2025_nes2net}, TCM-ADD~\cite{truong2024_tcmadd}); (ii) \textit{Graph-attention models for spectro-temporal relations} (AASIST3~\cite{aasist3_2024_borodin}); and (iii) \textit{SSL-based and large-scale pretrained detectors} (wav2vec~2.0-based deepfake detection~\cite{ssl_wav2vec2_tak2022}, SLS classifier with self-supervised XLS-R representations~\cite{zhang2024_sls_mm}, Arena audio deepfake detection checkpoints (500M and 1B variants)~\cite{kulkarni2025_df_arena_500m_hf,kulkarni2025_df_arena_1b_hf}). 

\subsection{Implementation and Inference Protocol}
All models were evaluated using their official open-source implementations or released checkpoints in a standardized inference pipeline, strictly adhering to the configurations provided in the respective evaluation scripts. Score normalization or calibration was not applied; raw output scores were used for all metric calculations.

\textbf{1) Pre-processing:}
All audio recordings were resampled to \textbf{16\,kHz} mono prior to inference. This sampling rate was consistent across all baselines, matching the training configuration of the provided checkpoints (e.g., ASVspoof-pretrained models).

\textbf{2) Fixed-Length Models (AASIST3, Res2TCNGuard, ResCapsGuard, TCM-ADD, Wav2Vec 2.0 anti-spoofing, SLS with XLS-R):}
Based on the provided inference protocols, these models were evaluated using a strict fixed-length input strategy to match their training constraints (typically ASVspoof LA/DF protocols).
\begin{itemize}
    \item \textit{Target Duration:} \textbf{64,600 samples} ($\approx$ 4.04\,s).
    \item \textit{Processing Logic:} A standardized padding function was applied: utterances shorter than the target length were repeated (looped) to fill the context window, while longer utterances were cropped to the first 64,600 samples.
    \item \textit{Score Extraction:} The processed fixed-length waveforms were fed into the models, and the final bona fide vs.~spoof score was taken directly from the output layer.
\end{itemize}

\textbf{3) Nes2Net:}
Evaluated using the official inference script in the \texttt{4s} mode.
\begin{itemize}
    \item \textit{Target Duration:} \textbf{64,000 samples} ($\approx$ 4.00\,s).
    \item \textit{Processing Logic:} Similar to the other baselines, inputs were padded via looping or cropped, but to a slightly shorter target duration defined by the Nes2Net architecture.
\end{itemize}

\textbf{4) Large-Scale Pretrained Detectors (Arena-1B/500M):}
Evaluated using the official Hugging Face \texttt{transformers} pipeline.
\begin{itemize}
    \item \textit{Target Duration:} \textbf{64,600 samples} ($\approx$ 4.04\,s).
    \item \textit{Inference:} The pipeline automatically handles feature extraction via the Conformer backbone, producing a sequence-level classification score for the entire recording.
\end{itemize}

\subsection{Results}
The baseline performance of all models on the \textbf{RuASD} clean test set is summarized in Table~\ref{tab:models_clean}, while the robustness of selected systems under codec- and channel-based augmentations is reported in Table~\ref{tab:models_aug}. Together, these two evaluations quantify in-domain discrimination on clean audio and robustness to realistic channel and platform effects under controlled degradations. When interpreting Table~\ref{tab:models_aug}, we compare models within the same augmentation condition (i.e., within a row), since different degradations induce different levels of task difficulty.

\section{Results}
\label{sec:results}

\begin{table*}
\centering
\caption{Antispoofing models on clean data}
\label{tab:models_clean}
\setlength{\tabcolsep}{2.5pt}

\begin{adjustbox}{width=\textwidth}

\begin{tabular}{l c c c c c c c}
\toprule
\textbf{Model} & {\textbf{Acc}} & {\textbf{Pr}} & {\textbf{Rec}} & {\textbf{F1}} & {\textbf{RAUC}} & {\textbf{EER}} & {\textbf{t-DCF}}\\
\midrule
AASIST3~\cite{aasist3_2024_borodin} & 0.769$\pm$0.0006 & 0.683$\pm$0.001 & 0.769$\pm$0.0006 & 0.724$\pm$0.001 & 0.841$\pm$0.0006 & 0.231$\pm$0.0006 & 0.702$\pm$0.002 \\
Arena-1B~\cite{kulkarni2025_df_arena_1b_hf} & \uline{0.812$\pm$0.001} & \uline{0.736$\pm$0.001} & \uline{0.812$\pm$0.001} & \uline{0.772$\pm$0.001} & \uline{0.887$\pm$0.0005} & \uline{0.188$\pm$0.001} & \textbf{0.385$\pm$0.001} \\
Arena-500M~\cite{kulkarni2025_df_arena_500m_hf} & 0.801$\pm$0.001 & 0.722$\pm$0.001 & 0.801$\pm$0.001 & 0.760$\pm$0.001 & 0.864$\pm$0.0005 & 0.199$\pm$0.001 & 0.655$\pm$0.002 \\
Nes2Net~\cite{liu2025_nes2net} & 0.689$\pm$0.0007 & 0.589$\pm$0.001 & 0.689$\pm$0.0007 & 0.634$\pm$0.0008 & 0.779$\pm$0.0007 & 0.311$\pm$0.0007 & 0.696$\pm$0.001 \\
Res2TCNGaurd~\cite{borodin2024_rescaps_res2tcn} & 0.627$\pm$0.001 & 0.520$\pm$0.001 & 0.627$\pm$0.001 & 0.569$\pm$0.001 & 0.691$\pm$0.001 & 0.373$\pm$0.001 & 0.918$\pm$0.001 \\
ResCapsGuard~\cite{borodin2024_rescaps_res2tcn} & 0.677$\pm$0.001 & 0.575$\pm$0.001 & 0.677$\pm$0.001 & 0.622$\pm$0.001 & 0.718$\pm$0.001 & 0.323$\pm$0.001 & 0.896$\pm$0.001 \\
SLS with XLS-R~\cite{zhang2024_sls_mm} & 0.779$\pm$0.001 & 0.700$\pm$0.001 & 0.779$\pm$0.001 & 0.737$\pm$0.001 & 0.859$\pm$0.001 & 0.221$\pm$0.001 & 0.650$\pm$0.001 \\
Wav2Vec 2.0~\cite{ssl_wav2vec2_tak2022} & 0.772$\pm$0.0006 & 0.687$\pm$0.001 & 0.772$\pm$0.0006 & 0.727$\pm$0.001 & 0.850$\pm$0.0006 & 0.228$\pm$0.0006 & 0.558$\pm$0.002 \\
TCM-ADD~\cite{truong2024_tcmadd} & \textbf{0.857$\pm$0.001} & \textbf{0.797$\pm$0.001} & \textbf{0.859$\pm$0.001} & \textbf{0.827$\pm$0.001} & \uline{0.914$\pm$0.0004} & \textbf{0.143$\pm$0.001} &  \textbf{0.424$\pm$0.001} \\
\bottomrule
\end{tabular}

\end{adjustbox}

\end{table*}
\begin{table}
\caption{Antispoofing models on augmented data (EER). \textbf{Aug.} denotes the applied degradation: \textbf{R}--RIR reverberation, \textbf{N}--MUSAN additive noise, and suffixes (\textit{alaw}, \textit{amr}, \textit{g722}, \textit{mp3}, \textit{mlw}, \textit{op16}, \textit{op8}, \textit{spx8}) indicate encode--decode transcoding with the corresponding codec; combined labels (e.g., \textbf{RNmp3}) apply \textbf{R}+\textbf{N} followed by codec transcoding.}
\label{tab:models_aug}

\centering
\setlength{\tabcolsep}{2.0pt}
\renewcommand{\arraystretch}{1.0}

\begin{adjustbox}{max width=0.48\textwidth}

\begin{tabular}{l c c c c c c c c c}
\toprule
\textbf{Aug.} & {\textbf{AAS3}} & {\textbf{AR1B}} & {\textbf{AR5M}} & {\textbf{N2NT}} &
{\textbf{R2NT}} & {\textbf{RCPS}} & {\textbf{XSLS}} & {\textbf{W2AS}} & {\textbf{TCM}} \\
\midrule

\multicolumn{10}{c}{\textbf{Codec only}} \\
\midrule
alaw  & 0.468 & 0.331 & \bfseries 0.237 & 0.435 & 0.331 & 0.332 & 0.485 & \uline{0.242} & 0.270 \\
amr   & 0.373 & \bfseries 0.133 & \uline{0.147} & 0.378 & 0.271 & 0.272 & 0.478 & 0.212 & 0.372 \\
g722  & 0.279 & 0.264 & \uline{0.245} & 0.353 & 0.323 & 0.323 & 0.473 & 0.275 & \bfseries 0.190 \\
mp3   & 0.239 & 0.191 & \bfseries 0.171 & 0.340 & 0.322 & 0.318 & 0.475 & 0.270 & \uline{0.187} \\
mlw   & 0.463 & 0.330 & \uline{0.238} & 0.449 & 0.333 & 0.325 & 0.478 & \bfseries 0.234 & 0.261 \\
op16  & 0.264 & \uline{0.216} & \bfseries 0.176 & 0.383 & 0.308 & 0.303 & 0.481 & 0.278 & 0.297 \\
op8   & 0.341 & \uline{0.208} & \bfseries 0.205 & 0.418 & 0.293 & 0.297 & 0.481 & 0.297 & 0.404 \\
spx8  & 0.372 & \bfseries 0.141 & \uline{0.147} & 0.370 & 0.305 & 0.302 & 0.475 & 0.273 & 0.420 \\
\midrule

\multicolumn{10}{c}{\textbf{Noise: N and N+Codec}} \\
\midrule
N     & 0.458 & 0.446 & 0.360 & 0.440 & 0.330 & \uline{0.310} & 0.481 & \bfseries 0.292 & \bfseries 0.292 \\
Nalaw & 0.503 & 0.430 & 0.321 & 0.483 & 0.320 & \uline{0.318} & 0.498 & \bfseries 0.291 & 0.380 \\
Namr  & 0.441 & \uline{0.264} & \bfseries 0.235 & 0.448 & 0.286 & 0.269 & 0.481 & 0.270 & 0.476 \\
Ng722 & 0.448 & 0.428 & 0.357 & 0.434 & 0.323 & \uline{0.305} & 0.479 & \bfseries 0.296 & \bfseries 0.296 \\
Nmp3  & 0.448 & 0.372 & \bfseries 0.291 & 0.437 & 0.327 & 0.312 & 0.476 & \uline{0.296} & \uline{0.296} \\
Nmlw  & 0.502 & 0.402 & \uline{0.304} & 0.480 & 0.319 & 0.316 & 0.497 & \bfseries 0.282 & 0.373 \\
Nop16 & 0.420 & 0.384 & \uline{0.305} & 0.447 & 0.321 & \bfseries 0.289 & 0.481 & 0.319 & 0.414 \\
Nop8  & 0.437 & 0.336 & 0.319 & 0.473 & \uline{0.309} & \bfseries 0.301 & 0.475 & 0.348 & 0.512 \\
Nspx8 & 0.430 & \uline{0.234} & \bfseries 0.217 & 0.428 & 0.288 & 0.273 & 0.479 & 0.312 & 0.490 \\
\midrule

\multicolumn{10}{c}{\textbf{Reverberation: R and R+Codec}} \\
\midrule
R     & 0.351 & 0.482 & \bfseries 0.319 & 0.499 & 0.332 & 0.336 & 0.483 & \uline{0.331} & 0.456 \\
Ralaw & 0.472 & 0.488 & 0.373 & 0.564 & \bfseries 0.313 & \uline{0.342} & 0.457 & 0.347 & 0.420 \\
Ramr  & 0.444 & 0.404 & \bfseries 0.288 & 0.515 & \uline{0.334} & 0.346 & 0.479 & 0.339 & 0.396 \\
Rg722 & 0.397 & 0.491 & \bfseries 0.305 & 0.500 & 0.337 & 0.348 & 0.489 & \uline{0.334} & 0.476 \\
Rmp3  & 0.394 & 0.444 & \bfseries 0.243 & 0.515 & \uline{0.326} & 0.336 & 0.491 & 0.336 & 0.454 \\
Rmlw  & 0.471 & 0.488 & 0.365 & 0.565 & \bfseries 0.318 & 0.346 & 0.468 & \uline{0.341} & 0.406 \\
Rop16 & 0.388 & 0.444 & \bfseries 0.295 & 0.489 & \uline{0.328} & 0.329 & 0.491 & 0.366 & 0.494 \\
Rop8  & 0.410 & 0.421 & \bfseries 0.308 & 0.509 & \uline{0.321} & 0.337 & 0.499 & 0.380 & 0.405 \\
Rspx8 & 0.454 & 0.400 & \bfseries 0.292 & 0.504 & \uline{0.316} & 0.341 & 0.490 & 0.361 & 0.413 \\
\midrule

\multicolumn{10}{c}{\textbf{Combined: RN and RN+Codec}} \\
\midrule
RN     & 0.503 & 0.486 & 0.408 & 0.527 & \bfseries 0.324 & 0.379 & 0.504 & \uline{0.365} & 0.511 \\
RNalaw & 0.493 & 0.493 & 0.401 & 0.551 & \bfseries 0.316 & 0.381 & 0.528 & \uline{0.372} & 0.422 \\
RNamr  & 0.473 & 0.478 & \uline{0.349} & 0.515 & \bfseries 0.319 & 0.377 & 0.512 & 0.352 & 0.385 \\
RNg722 & 0.477 & 0.498 & 0.401 & 0.529 & \bfseries 0.322 & 0.384 & 0.504 & \uline{0.366} & 0.473 \\
RNmp3  & 0.469 & 0.487 & \uline{0.350} & 0.530 & \bfseries 0.332 & 0.371 & 0.505 & 0.366 & 0.510 \\
RNmlw  & 0.503 & 0.492 & 0.397 & 0.544 & \bfseries 0.310 & \uline{0.377} & 0.510 & 0.379 & 0.409 \\
RNop16 & 0.477 & 0.498 & 0.403 & 0.519 & \bfseries 0.329 & \uline{0.385} & 0.506 & 0.400 & 0.427 \\
RNop8  & 0.493 & 0.496 & 0.405 & 0.510 & \bfseries 0.324 & 0.384 & 0.488 & 0.390 & \uline{0.379} \\
RNspx8 & 0.477 & 0.449 & \uline{0.333} & 0.507 & \bfseries 0.318 & 0.387 & 0.497 & 0.379 & 0.391 \\
\bottomrule
\end{tabular}

\end{adjustbox}

\end{table}

This section reports benchmark results for the evaluated anti-spoofing models on \textbf{RuASD}.
Performance is assessed under two conditions: (i) \textit{clean} (no artificial degradations) and (ii) \textit{augmented} (controlled degradations emulating realistic distribution channels and post-processing, including codec transcoding, additive noise from MUSAN, and room reverberation using RIRs).
This evaluation setting targets robustness under domain shifts, which is consistent with modern ASVspoof-style protocols and related benchmarks \cite{asvspoof2021_accelerating,asvspoof2021_summary,asvspoof2019_overview,muller2022_in_the_wild}.
Detection quality is quantified using Accuracy (Acc), Precision (Pr), Recall (Rec), F1-score, ROC-AUC (denoted as RAUC), and Equal Error Rate (EER), where EER is defined as the error rate at the operating point where the False Acceptance Rate (FAR) equals the False Rejection Rate (FRR) \cite{asvspoof2021_summary}.

Overall, the benchmark remains challenging and indicates substantial headroom for improvement.
Even on \textit{clean} data, the best-performing systems do not reach perfect discrimination (RAUC $\le$ 0.891) and exhibit non-negligible EER values (down to 0.159) (Table~\ref{tab:models_clean}).
Under realistic channel and post-processing effects, performance degrades across all evaluated model families (Table~\ref{tab:models_aug}), underscoring the need for robustness-oriented development.

\subsection{Results on Clean Data}
\label{subsec:results_clean}

Table~\ref{tab:models_clean} summarizes model performance on the \textit{clean} test set. 
The best overall results are obtained by TCM-ADD (Acc=0.857, RAUC=0.914, EER=0.143) and the large-capacity Arena detectors (Arena-1B: Acc=0.812, RAUC=0.887, EER=0.188; Arena-500M: Acc=0.801, RAUC=0.864, EER=0.199).
Among compact supervised backbones, AASIST3 provides competitive performance (Acc=0.769, RAUC=0.841, EER=0.231), whereas Nes2Net, Res2TCNGuard, and ResCapsGuard yield lower clean-set performance (Table~\ref{tab:models_clean}).
SSL-based baselines (SLS with XLS-R and Wav2Vec 2.0 anti-spoofing) achieve mid-to-high performance but remain below the best systems on this protocol (Table~\ref{tab:models_clean}).

\subsection{Robustness to Degradations}
\label{subsec:results_aug}

Table~\ref{tab:models_aug} reports EER under four augmentation subgroups: (i) \textbf{codec-only} transcoding, (ii) \textbf{additive noise} (\textbf{N}) and \textbf{N+codec}, (iii) \textbf{reverberation} (\textbf{R}) and \textbf{R+codec}, and (iv) \textbf{combined} \textbf{RN} and \textbf{RN+codec}.

\paragraph{Codec-only.}
In codec-only conditions, the best EERs are achieved by Arena-1B/Arena-500M, Wav2Vec 2.0 anti-spoofing, or TCM-ADD depending on the codec (Table~\ref{tab:models_aug}).
Arena-1B yields the lowest EER on \textit{amr} and \textit{spx8} (0.133 and 0.141), while Arena-500M is best on \textit{alaw}, \textit{mp3} \textit{op8} and \textit{op16} (0.237, 0.171, 0.176 and 0.205).
TCM-ADD is best on \textit{g722} (0.190), and Wav2Vec 2.0 anti-spoofing attains the lowest EER on \textit{mlw} (0.234) (Table~\ref{tab:models_aug}).

\paragraph{Additive noise (N) and N+codec.}
On \textbf{N}, the lowest EER is shared by Wav2Vec 2.0 anti-spoofing and TCM-ADD (both 0.292), followed by ResCapsGuard (0.311), while Arena-500M and Arena-1B yield EER=0.360 and EER=0.446, respectively (Table~\ref{tab:models_aug}).
For N+codec settings, Arena-500M achieves the best EER on \textit{Namr}, \textit{Nmp3} and \textit{Nspx8} (0.235, 0.291 and 0.217), whereas Wav2Vec 2.0 anti-spoofing and ResCapsGuard yield the best EER for several other conditions (e.g., \textit{Nalaw}: 0.291; \textit{Nmlw}: 0.282; \textit{Nop16}: 0.289; \textit{Nop8}: 0.301) (Table~\ref{tab:models_aug}).

\paragraph{Reverberation (R) and R+codec.}
For reverberation alone (\textbf{R}), the best EER is obtained by Arena-500M (0.319), followed by Wav2Vec 2.0 anti-spoofing (0.331) and Res2TCNGuard (0.332), while Arena-1B yields EER=0.482 (Table~\ref{tab:models_aug}).
For \textbf{R+codec} settings, the best-performing model depends on the codec: \textit{Res2TCNGuard} is strongest for \textit{Ralaw} (0.313) and \textit{Rmlw} (0.318), whereas \textit{Arena-500M} achieves the lowest EER for the remaining listed codecs (e.g., \textit{Ramr}: 0.288; \textit{Rmp3}: 0.243; \textit{Rspx8}: 0.292) (Table~\ref{tab:models_aug}).

\paragraph{Combined degradations (RN) and RN+codec.}
Under \textbf{RN}, the lowest EER is obtained by Res2TCNGuard (0.324), followed by Wav2Vec 2.0 anti-spoofing (0.365) (Table~\ref{tab:models_aug}).
For \textbf{RN+codec} settings, Res2TCNGuard yields the lowest EER for every codec (\textit{RNalaw}: 0.316; \textit{RNmlw}: 0.310; \textit{RNspx8}: 0.318), and Arena-500M and Wav2Vec 2.0 anti-spoofing are typically the next-lowest (Table~\ref{tab:models_aug}).
Across RN/RN+codec settings, EERs are higher than in many single-factor conditions for the same models (Table~\ref{tab:models_aug}).

\section{Discussion}
The presented corpus targets a realistic Russian-language anti-spoofing setting in which both the attack surface (diverse modern TTS generators) and the dissemination channel (noise, reverberation, and transcoding) introduce substantial distribution shift relative to conventional clean-data protocols. In this context, the results highlight three practically relevant observations: (i) high perceptual quality of spoofed speech does not imply easy detectability, (ii) robustness to channel and codec effects is a first-order requirement, and (iii) model families differ in their sensitivity to protocol choices and post-processing.

\subsection{Spoof realism and heterogeneity}
The objective descriptors and distributions indicate that the spoof subset contains a wide spectrum of attack quality rather than a single homogeneous generator regime. In particular, MOS and CER vary substantially across TTS systems (Table~\ref{tab:models_clean} and Figs.~\ref{fig:spoof_mos}--\ref{fig:spoof_cer}), including generators with narrow high-MOS distributions close to the upper end of the scale as well as pipelines with clearly lower MOS and/or higher CER. This heterogeneity is important for evaluation because it avoids a ``trivial'' corpus dominated either by obviously degraded synthetic speech or by a single strong generator family. The bona fide subset is likewise non-uniform: MOS and CER distributions differ across source corpora (Table~\ref{tab:models_aug}, Fig.~\ref{fig:bonafide_descriptors}(a) and Fig.~\ref{fig:bonafide_descriptors}(b)), reflecting diverse recording conditions and annotation quality typical for found data, while OpenSTT exhibits a high MOS but lacks reference transcripts for CER computation (Table~\ref{tab:bonafide_dataset_stats}). Overall, these statistics support that the proposed corpus covers both realistic attacks and realistic bona fide variability, motivating evaluation beyond a single clean-domain operating point.

\subsection{Clean-data accuracy vs.\ robustness under degradations}
On clean data (Table~\ref{tab:models_clean}), TCM-ADD achieves the best overall performance across all reported metrics (Accuracy/F1/ROC-AUC) and also attains the lowest EER, indicating the strongest discrimination under the clean protocol.
The Arena models follow closely in clean-set ranking (with Arena-1B outperforming Arena-500M), while AASIST3, SLS with XLS-R, and Wav2Vec 2.0 anti-spoofing form a competitive mid-tier; in contrast, lightweight supervised backbones trail in clean accuracy and exhibit higher EER values.
However, the augmented evaluation (Table~\ref{tab:models_aug}) shows that clean-set superiority does not directly translate to robustness: performance can degrade sharply under channel shift, especially when multiple distortion factors are combined (RN and RN+codec).
In particular, additive noise (N) and reverberation (R) already increase EER substantially for several top clean-set systems, and the RN family of conditions typically yields among the largest EER values, emphasizing that robustness must be treated as a primary evaluation axis for deployment-facing countermeasures.

\subsection{Model family behavior and protocol sensitivity}
The robustness patterns in Table~\ref{tab:models_aug} indicate a clear trade-off between clean-set accuracy and stability under severe distortions, and the relative ranking can change substantially as the degradation regime shifts from codec-only to channel-impaired conditions (additive noise and/or reverberation), including their combined RN variants with optional transcoding.
For \textit{codec-only} conditions, the Arena detectors provide strong robustness overall: Arena-500M achieves the lowest EER for several codecs (e.g., \textit{alaw}, \textit{mp3}, \textit{op16} and \textit{op8}), while Arena-1B is strongest for some narrowband-style codecs (e.g., \textit{amr} and \textit{spx8}), indicating heterogeneous codec sensitivity even within the same model family.
AASIST3 remains competitive on wideband codecs (e.g., \textit{mp3}/\textit{op16}) but degrades sharply under narrowband-like transcodings such as \textit{alaw} and \textit{mlw}, consistent with codec-induced spectral modifications shifting the operating point even when the architecture captures informative spectro-temporal artifacts.

For \textit{noise-driven} conditions (\textbf{N} and \textbf{N}+codec), additive noise increases EER for all evaluated model families, and the relative robustness depends on the specific codec applied on top of noise.
The Arena models remain comparatively strong for several N+codec settings (e.g., \textbf{Namr}, \textbf{Nmp3} and \textbf{Nspx8}), whereas TCM-ADD and SSL-based baselines exhibit larger EER increases in some noisy-transcoded regimes (e.g., \textbf{Nop8}), indicating that clean-set ranking does not reliably predict performance under combined distortions.
Lightweight supervised models (Res2TCNGuard/ResCapsGuard) do not yield the lowest EER under \textbf{N} alone, but their EER varies less across the \textbf{N} and \textbf{N}+codec conditions, suggesting more consistent behavior under additive interference.

For \textit{reverberation-driven} conditions (\textbf{R} and \textbf{R}+codec), reverberation induces a degradation pattern that differs from additive noise, and the addition of transcoding can further increase operating-point errors. 
Arena-500M shows a substantial EER increase under \textbf{R} relative to codec-only conditions, however, it still delivers the lowest EER among the evaluated models. The lightweight supervised backbones remain competitive in several R+codec settings (e.g., \textbf{Ralaw} and \textbf{Rmlw}), indicating that clean-set ranking does not reliably predict performance under reverberant and transcoded channels.
Overall, the \textbf{R} and \textbf{R}+codec results show that robustness to room acoustics should be evaluated explicitly rather than inferred from noise-only or codec-only performance.

In the combined degradation conditions (\textbf{RN} and \textbf{RN}+codec), the relative ranking changes: Res2TCNGuard achieves the lowest EER for all RN conditions, despite comparatively weak clean evaluation condition performance. 
This inversion indicates that robustness under multi-factor channel shift (reverberation and additive noise, optionally followed by transcoding) is not implied by clean-set ranking, and it motivates reporting both clean and augmented results when assessing deployment behavior. These findings underscore that robustness is highly condition-specific, motivating a closer look at the evaluation scope and factors that may limit the generality of the observed rankings.

\subsection{Limitations and implications for future datasets}
While MOS/CER provide useful dataset descriptors, they are imperfect proxies for adversarial strength: high MOS or low CER can coexist with detectable artifacts, and conversely, low-quality samples may be easy to detect but less representative of real attacks. Moreover, CER availability depends on reference transcripts and is therefore missing for some sources (e.g., OpenSTT in Table~\ref{tab:bonafide_dataset_stats} and some TTS systems in Table~\ref{tab:spoof_model_stats}), which limits direct subset-level intelligibility comparison without additional transcription alignment or manual verification. Finally, the fixed-length evaluation used by several baselines may under-utilize long-range cues present in longer utterances and can interact with codec/noise effects; reporting both fixed-length and variable-length inference, when supported by the model, would further clarify deployment behavior.

Beyond metric availability, the corpus inherits several dataset-design constraints. First, spoofed utterances are synthesized from texts drawn from a single domain, which may limit lexical and stylistic diversity and can make the evaluation less representative of real-world attacks that span broader content distributions. Second, our spoof set focuses on fully synthetic utterances and does not include partially manipulated samples such as local word/phrase replacement or splicing/montage edits; detectors that rely on global utterance-level cues may therefore be overestimated relative to scenarios where only short regions are tampered with. Third, the bona fide portion is pooled from multiple Russian ASR corpora with heterogeneous domains, recording channels, and transcription quality; this heterogeneity can introduce source-specific ``fingerprints'' that models may inadvertently exploit, partially learning to separate dataset sources rather than bona fide versus spoof artifacts.

These limitations suggest that future extensions should (i) expand transcript coverage for bona fide sources where feasible, (ii) broaden text-domain coverage for spoof generation and incorporate partially edited manipulations, and (iii) extend the augmentation suite with platform- and device-driven post-processing chains observed in real dissemination to reduce the gap between controlled degradations and in-the-wild audio.

\section{Conclusion}
\label{sec:conclusion}
This paper introduced \textbf{RuASD} (Russian AntiSpoofing Dataset), a dedicated, reproducible Russian-language benchmark for speech anti-spoofing that targets modern threat models combining diverse spoof generators with deployment-inspired channel and platform distortions.
RuASD includes spoofed speech synthesized using 37 Russian-capable TTS systems and a bona fide subset curated from heterogeneous Russian speech corpora, enabling systematic evaluation beyond a single clean-domain operating point.

We provided a unified simulation framework covering reverberation, additive noise, and a range of speech-codec transcodings, treating these conditions as controlled proxies for real dissemination pipelines.
Using this protocol, we benchmarked a representative set of publicly available countermeasures spanning compact supervised backbones, graph-attention models, SSL-based detectors, and large-scale pretrained systems, and we reported reference results on both clean and augmented conditions.

The obtained baselines confirm that the benchmark remains challenging: even on clean audio, the best-performing system reaches ROC-AUC 0.891 with EER 0.159, while performance deteriorates substantially under realistic distortions, especially for combined noise-and-reverberation conditions.
Importantly, we observe that clean-set ranking does not reliably predict robustness under channel shift, motivating robustness-oriented reporting as a first-class requirement for deployment-facing anti-spoofing evaluation.

Overall, \textbf{RuASD} fills a gap in Russian-language resources and provides a reproducible testbed for developing and comparing anti-spoofing systems under realistic distribution shift.
Future work will focus on expanding content diversity, increasing transcript coverage where feasible, and extending the attack set toward partially manipulated and in-the-wild scenarios to further reduce the gap between controlled evaluation and real-world threats.

\bibliographystyle{IEEEtran}
\bibliography{sections/references}

\begin{IEEEbiography}[{\includegraphics[width=1in,height=1.25in,clip,keepaspectratio]{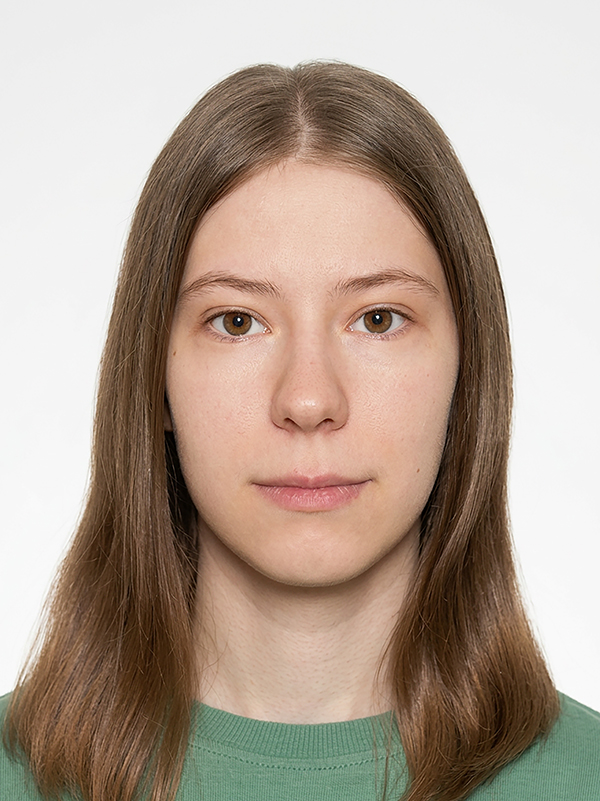}}]{Ksenia Lysikova}
was born in Moscow, Russia, in 2004. She is currently pursuing the B.S. degree in computer science and computer engineering at the Moscow Technical University of Communications and Informatics (MTUCI), Moscow, Russia.

She is a Researcher in the "Intelligent Systems" Research Department, MTUCI, Moscow, Russia. Her research focuses on speech technologies and anti-spoofing methods. Her research interests include speech anti-spoofing, ASR models, NLP/LLM systems,  Neural networks interpretability, and AI safety.

Ms. Lysikova is not a member of any professional societies.
\end{IEEEbiography}

\begin{IEEEbiography}[{\includegraphics[width=1in,height=1.25in,clip,keepaspectratio]{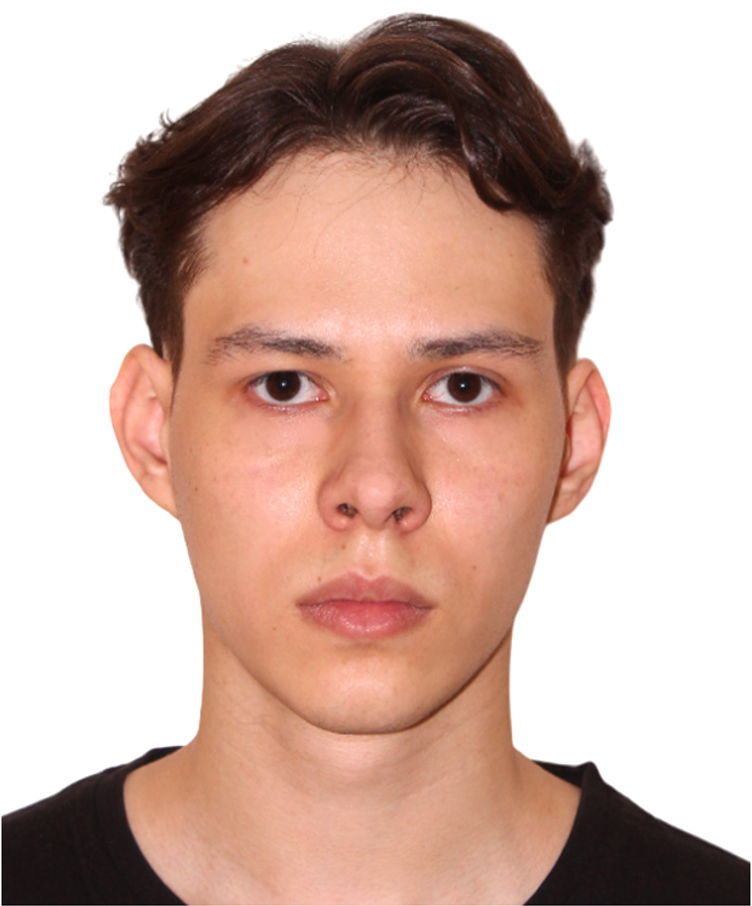}}]{Kirill Borodin} was born in Volgograd, Russia, in 2004. He is currently completing the B.S. degree in computer science with the Moscow Technical University of Communications and Informatics (MTUCI), Moscow, Russia, expected in 2026.

He is a Researcher in the "Intelligent Systems" Research department, MTUCI, Moscow, Russia. His research interests include speech anti-spoofing, large-scale data filtration and annotation, diffusion and flow matching speech synthesis, and representation learning. He is a first author of "AASIST3: KAN-enhanced AASIST speech deepfake detection using SSL features and additional regularization for the ASVspoof 2024 Challenge" (Proc. The Automatic Speaker Verification Spoofing Countermeasures Workshop (ASVspoof 2024), 2024, pp. 48--55).

Mr. Borodin is not a member of any professional societies.
\end{IEEEbiography}

\begin{IEEEbiography}[{\includegraphics[width=1in,height=1.25in,clip,keepaspectratio]{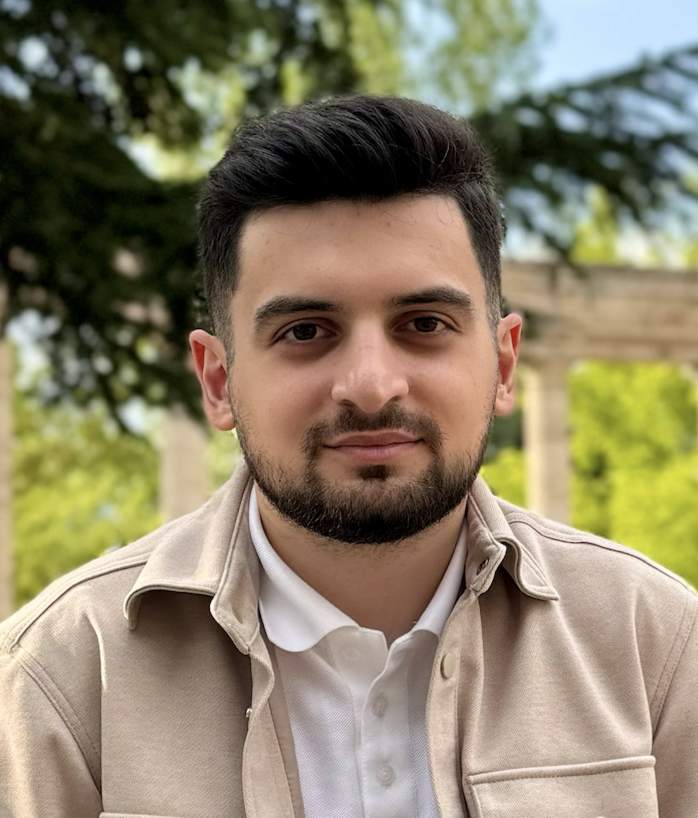}}]{Grach Mkrtchian} received the M.Sc. degree in 2021 from the Moscow Technical University of Communications and Informatics (MTUCI). He also holds a Ph.D. degree in Computer Science.

He is currently the Head of the Research Department “Intelligent Systems,” where he leads research and development in applied artificial intelligence, intelligent sensing, and data-driven system design. His work focuses on AI technologies with measurable practical impact, including signal processing, computer vision, multimodal data analysis, and robust perception systems for complex real-world environments.

His research interests include end-to-end AI system design, integration of machine learning with engineering systems, evaluation methodologies, system reliability, and performance validation under operational conditions.
\end{IEEEbiography}

\EOD

\end{document}